TITLE

An in-vivo study of electrical charge distribution on the bacterial cell wall by Atomic Force Microscopy in vibrating force mode.

AUTHORS

Christian Marlière[1]* and Samia Dhahri[2]

AFFILIATIONS

1. Institut des Sciences Moléculaires d'Orsay, ISMO, University Paris-Sud, CNRS, Orsay, France.
2. University of Carthage, High Institute of Environmental Science and Technology of Borj Cedria, Ben Arous, Tunisia.

* CORRESPONDING AUTHOR

**Dr. Christian Marlière**
Institut des Sciences Moléculaires d'Orsay (ISMO), Université Paris-Sud, CNRS,
Bâtiment 350, Université Paris-Sud, 91405 Orsay Cedex, France,
Phone: (+33) 169 157 511 / email: christian.marliere@u-psud.fr

## ABSTRACT

We report an in-vivo electromechanical Atomic Force Microscopy (AFM) study of charge distribution on the cell wall of Gram+ *Rhodococcus wratislaviensis* bacteria, naturally adherent to a glass substrate, in physiological conditions. The method presented in this paper relies on a detailed study of AFM approach/retract curves giving the variation of the interaction force versus distance between tip and sample. In addition to classical height and mechanical (as stiffness) data, mapping of local electrical properties, as bacterial surface charge, was proved to be feasible at a spatial resolution better than few tens of nanometers. This innovative method relies on the measurement of the cantilever's surface stress through its deflection far from (>10nm) the repulsive contact zone: the variations of surface stress come from modification of electrical surface charge of the cantilever (as in classical electrocapillary measurements) likely stemming from its charging during contact of both tip and sample electrical double layers. This method offers an important improvement in local electrical and electrochemical measurements at the solid/liquid interface particularly in high-molarity electrolytes when compared to technics focused on the direct use of electrostatic force. It thus opens a new way to directly investigate *in-situ* biological electrical surface processes involved in numerous practical and fundamental problems as bacterial adhesion, biofilm formation, microbial fuel cell, etc.

# Table of contents

In-vivo AFM study of charge distribution on bacterial cell wall is proved to be feasible at a nanometric spatial.

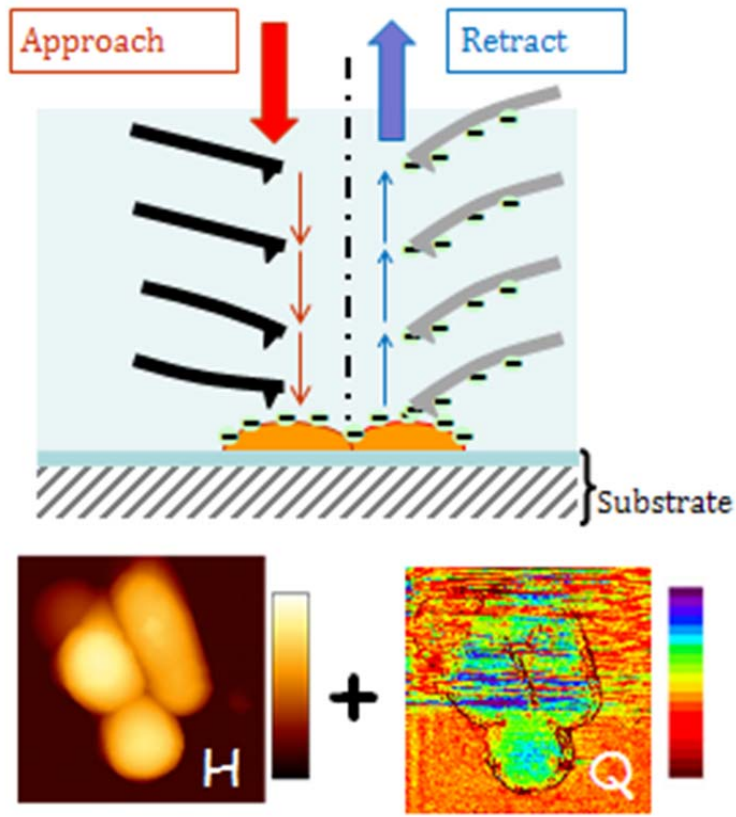

## INTRODUCTION

Solid/liquid interfaces may be subject to bacterial adhesion and biofilm formation. If the microbial consortia are pathogenic these surfaces will be a starting point for nosocomial and food born infections. It is well known that bacterial adhesion on inert surfaces is mainly steered by non-covalent molecular interactions with long-range interactions as van der Waals, electrostatic (mainly resulting from the overlapping of electrical double layers) or short-range ones as Lewis acid-base and hydration interactions[1]. However the knowledge of basic processes governing the bacterial adhesion to abiotic surfaces still remains an important question. Electrostatic interactions is one of the key factor as the bacterial cell surface is known to carry a net negative charge under most of physiological conditions[2]. The problem is intricate as bacterial surfaces are chemically and structurally heterogeneous[3]. As an example, the bacterial membrane contents pore-forming integral membrane proteins constituting channels of conduction for charged particles. Ions, typically, will thus flow through the cellular membrane provided that a driving force exists for ionic movement as, for instance electrochemical gradient[4]. Studies of these local electrophysiological properties of bacteria have been mainly performed by the patch-clamp technique. For it, a small patch of membrane containing one or a few channels (few tenths of nanometers in diameter) needs to be isolated and placed at the tip of a micropipette (tip opening ~1 μm), filled with an ionic solution. The conditioning of the membrane's sample for patch-clamp experiments resorts to aggressive technics as lysozyme digestion, a mandatory step to the completion of needed spheroplasts. Thanks to this technics important results on the gating kinetics and ion permeability of membrane channels sensitive to osmotic pressure were obtained[5,6].

However few are known about electrical properties of the bacterial cell surface at the nanometer level and in low perturbative conditions as valuable methods for investigating local surface charge or potential distribution are still missing especially in aqueous solutions with high ionic strength. Atomic Force Microscopy (AFM) may be a valuable choice for local Investigation of dielectric properties in physiological environments as nanometric spatial resolution is easily reached and AFM can be used in various electrical or electrochemical modes. Significant progress has been made with AFM operating in liquid environments when focusing on topographic[7–9] or mechanical[10] aspects. . In ambient and vacuum environments, the electrostatic properties of surfaces can be easily mapped by Kelvin probe force microscopy[11]. The presence of mobile ions in liquid environment complicates the implementation of such technics so that it was mainly applied to non-polar liquids (i.e. containing few mobile ions)[12] or in low-molarity electrolytes (<10mM)[13,14]. Presence of mobile ions and bias voltage applied between tip and sample lead to induced charge dynamics, ion diffusion and a capacitive coupling between them depending on the electrochemical properties of the solution (ionic strength, etc.)[15]. In search of quantitative imaging of local dielectric properties in electrolyte solutions with nanoscale spatial resolution, most of studies have been done by applying high frequency voltages[16] between the cantilever and sample. By working at frequencies greater than the dielectric relaxation frequency of the medium (in the MHz range for 1mM KCl[17]) and by scanning the probe at distances much larger than the Debye length (typically 10 nm in 1mM monovalent electrolytes), spatially resolved mapping of the dielectric properties of patterned samples are feasible[18]. Working at high frequency is an efficient way to actuate the cantilever solely by electrostatic force and not by surface stress forces that contribute significantly to the cantilever vibration in the low frequency regime[17,19]. Derived methods based on similar concept were used to characterize samples with charged domains in liquid media at low ionic strength[13,14,20]. As voltage varies at time scale much lower than typical time related to bulk diffusion of ions (few microseconds for a distance of 100nm[21]) these technics avoid the formation of the electrical double layer (EDL) and hence hinders its use to the study of the EDL itself and consequently of the underlying surface repartition of charges on sample.

In this paper we present a new experimental method for investigation of local electrical surface charges. It is based on the original combination of two complementary physical processes already largely documented in literature. The first one is based on the study of the electrical double layer (EDL) by the use of immersion and emersion of a metallic electrode. From a pioneering idea of Kenrick[22], this method was used to study different properties of the EDL: potential of zero charge [23,24], ion absorption [25] etc. It was shown that the electrode can be removed from an electrolyte to air[26] or vacuum[25] as well. It was shown that the electric double layer is then intact, providing that there is no faradic current at the used voltage. It was thus proved that both the charge on the electrode and the potential across the emerging double layer remain fixed as this electrode is removed[27]. The proposed model [25] was based on the hypothesis the "unzipping" of the EDL during electrode's emersion takes place just outside the outer Helmholtz plane as defined in classical theory of EDL[28,29]. Probing of the EDL by a metallic probe was recently revisited at nanometer scale by Yoon *et al*[30].

The second physical processes involved in the therein presented method is based on the so-called electrocapillary equation (numbered 13.1.31 in reference [28]), relating changes in (i) the surface (or interfacial) tension, (ii) the electrode potential, (iii) its surface charge density, (iv) the electrolytic solution composition (by means of the electrochemical potentials) and finally (v) relative surface excesses of ions. This equation theoretically stemmed from the thermodynamical developing of Gibbs adsorption isotherm. It permitted the interpretation of surface tension measurements at mercury-electrolyte interfaces in the dropping mercury electrode (DME) method and consequently substantial improvement in knowledge on EDL structure. An elegant way of measuring, at sub-micrometric scale, the surface tension is to study variations of the surface stress of AFM cantilevers as these two quantities are linked thanks to Shuttleworth equation[31]. Variations in the surface stress indeed causes the bending of the AFM cantilever the amplitude of which can be measured thanks to the reflection of the laser beam as currently used in AFM experiments[32]. Several authors took advantage of that concept to measure adsorbate-induced surface stress changes in vacuum[33], electrocapillary-like curves of noble metals[34], changes of the surface stress of silicon nitride upon varying the pH[35], binding of proteins in aqueous electrolyte[36] or pH-induced protein conformation changes[37]. As variation of electrical properties (charge, potential) of the thin plate generates changes in the surface tension[28,38] via the electrocapillary equation, we decided to investigate local electrical properties of substrates of biological interest by AFM in an approach/retract mode at high frequency (≈ 0.1kHz). These experiments were done in aqueous liquid phase with a "high" (≈0.15M) ionic force (Debye length, $\lambda_D$, in the range of tenths of nanometers), i.e. in physiological conditions. The approach/retract movement of the AFM tip at every pixel could be split in two main segments: the first one corresponds to a zone of "high" distance from the substrate (it means larger than several $\lambda_D$) where electrical state of the cantilever's EDL is constant and a second one where the two EDL are in contact. In this last one the AFM tip's EDL reaches a new state of equilibrium leading to a change in surface stress: this is detected by the change of flexion of the cantilever as measured by the "force" signal far away ($\gg \lambda_D$) off the substrate. We studied living bacteria in their standard physiological conditions and naturally adherent (i.e. without any forced immobilization) to the glass substrate. We proved that, provided the adequate choice of experimental parameters (mainly time constants), imaging of local bacterial surface density of charge in its steady state is feasible at a spatial resolution better than few tens of nanometers. Furthermore, a dynamic effect of electrical charging was detected both by the new electro-mechanical method we present in this paper and by direct current measurement as well. This last effect was tentatively attributed to the detection of ionic current stemming from bacterial membrane pores.

## MATERIAL AND METHODS

**Bacterial preparation**

This study was done with *Rhodococcus wratislaviensis* (*Rhodo. w.*) bacterial strain known for its ability to degrade hydrocarbon compounds in aqueous effluents. This strain is registered at the Collection Nationale de Cultures de Microorganismes (CNCM), Paris, France under number CNCM I-4088 and was provided to us by IFPEN). Stock cultures were kept frozen at -80 °C in 20% glycerol (v/v). The culture medium used was a vitamin-supplemented mineral medium (MM). This medium contained $KH_2PO_4$, 1.40 g.l$^{-1}$; $K_2HPO_4$, 1.70 g.l$^{-1}$; $MgSO_4$ 7 $H_2O$, 0.5 g.l$^{-1}$; $NH_4NO_3$, 1.5 g.l$^{-1}$; $CaCl_2$ 2 $H_2O$, 0.04 g.l$^{-1}$. A vitamin solution and an oligo-element solution were added as previously described [39,40]. The pH of this medium is equal to 6.9[40]. After inoculation (10%), the adequate carbon source (in the present case toluene) was added, and the cultures were incubated at 30°C with constant agitation. Cultures were grown in flasks closed with a cap equipped with an internal Teflon septum to avoid any loss of substrate either by volatilization or by adsorption. The headspace volume was sufficient to prevent any $O_2$ limitation during growth. Growth was followed by measuring the optical density at a wavelength of 660nm. Bacteria were transplanted in fresh medium once a week. 500µL of this culture medium was pipetted when strain was in its exponential growth phase and introduced in the AFM liquid cell. It must be emphasized that we proved[10] that bacteria studied in these conditions are alive.

**AFM sample preparation**

The samples we used for the AFM experiments were soda-lime glass substrates recovered by an indium-tin-oxide (ITO) layer (purchased from Neyco, Paris, France). These samples were then cleaned by sonication in a diluted solution of detergent (pH around 9) for 15 minutes before being carefully rinsed with high purity water (Milli-Q). Drying was done below the flux of a pure inert gas.

The sample's preparation was derived from that extensively described in [10]. To resume: the bacterial suspension in its culture medium was gently vortexed during three minutes. Forty microliters (µL) were then deposited and remained on the glass slide during 10mn. The excess of solution was thereafter removed and the substrate was further left in surrounding atmosphere (22°C and around 60% of relative humidity) till the dehydration front started moving throughout the sample (around 5 minutes). The sample was then rinsed twice with the culture medium in gentle conditions before being placed at the bottom of the liquid cell, ECCell® from JPK [41]. Finally 500µL of the MM medium were promptly poured in the liquid cell. The final bacterial surface concentration on the glass substrate for the AFM experiments was around $2.10^5$ units per mm$^2$, as checked by optical microscopy. No spontaneous detachment of bacteria from the sample towards the planktonic phase was evidenced by optical or AFM microscopy. This method was proved [10] to be very efficient for AFM imaging of alive bacteria in their genuine physiological conditions. No aggressive external immobilization protocol, neither chemical nor mechanical, was needed: *Rhodo w.* self-immobilized on the ITO/glass substrate. In identical conditions gliding movements of cyanobacteria was studied in real time by the AFM tip[10] directly proving its minute level of disturbances and of course its non-lethal characteristics. These *Rhodo. w.* bacteria, in the initial stage of biofilm forming, are then studied by AFM in their genuine physiological condition. As no chemical templates for immobilization (as gelatin, polylysine etc) are used, bacterial membrane is not recovered by any polymeric exogenic compounds: the present electrical measurements are directly related to phenomena inherent to bacterial membrane eventually through a self-secreted extra polymeric substance (EPS).

**AFM data acquisition:**

Atomic force microscopy studies were carried out using a Nanowizard III (JPK Instruments AG, Berlin, Germany) and its electrochemical cell (ECCell® from JPK [41]). The AFM head is working on a commercial inverted microscope (Axio Observer.Z1, Carl Zeiss, Göttingen, Germany). This combined AFM/optical microscope was placed on an isolation vibration table. All the experiments were done with *Rhodo w.* in their liquid culture medium at a temperature of 24.0±0.2°C. It must be emphasized that a special care was put on thermal stability in order to minimize temperature fluctuation: particularly the experimental setup was placed in a temperature-controlled room located in the basement floor to minimize both building vibrations and thermal drift. As detailed in results section, residual thermal drift was measured and corrected as it was found to be linear with time.

AFM measurements were done using a fast-speed approach/retract mode (Quantitative Imaging® (QI) mode). At each pixel of the image, a complete force-distance curve, at a defined constant velocity, is acquired. In all presented results, the pixel-by-pixel extend/retract curves were done at a constant speed of 125µm/s on a total extension of 600nm (data were stored for the first 500nm from the substrate during both approach and retract movements of the cantilever). Thus the cantilever oscillating frequency is near to 85Hz. Two hundred points were acquired during each approach or retract curves. Typical images were done on the basis of a surface scanning with 64 by 64 or 128 by 128 pixels. We used standard beam AFM probes (PPP-CONTPt, Nanosensors, Neuchatel, Switzerland) with a nominal value of stiffness around 0.30N/m. Their values were precisely measured by thermal noise[42]. The sensitivity of detection of the vertical deflection thanks to the photodiode system was measured during the approach to a clean glass substrate. The typical tip height is about 15 microns. These commercial cantilevers are coated by a 25 nm thick double layer of chromium and platinum-iridium alloy on both sides. The maximum applied force was set at 6nN. No major changes in the quality of AFM data as presented here were observed with time. No contamination of the apex of the tip was detectable during experiments. Each AFM image is scanned line by line, starting from the bottom of the image to its top. For each of these lines pixels are successively scanned from the left side to the right one.

All the presented height AFM images and approach/retract curves are raw data (without any post-treatment as flattening, filtering, smoothing, etc.). Mathematical treatments, as calculation of the so-called Basic_Line Force signal (BL_Force; see below) or stiffness data were done by custom Matlab (from MathWorks, Natick, USA) programs. The stiffness data were calculated from the slope of the approach (i.e. extend) curves (force versus scanner elongation) at point of maximum force as averaged on a distance interval of 10nm. Other numerical treatments of AFM data were done using OriginPro software (from OriginLab Corporation, Northampton, USA).

**Electrical/electro-mechanical measurements.**

The presented AFM experiments were done in the JPK's electrochemical cell equipped with three-electrodes. The first one, the working electrode (WE), was the Pt-covered AFM cantilever; the counter-electrode (CE) is the ITO/glass plate. The third electrode is a platinum wire (0.6mm in diameter) with a ring-shape: its distance to the AFM tip (and consequently to the CE) is equal to ≈ 8mm; its diameter is equal to 15µm (see figure 1.a for a schematic). This was the reference electrode (REF) from which the electrical potentials were measured. It must be emphasized that we preferred not to use the classical Ag/AgCl or calomel quasi reference electrodes in order to avoid any contamination of the culture medium by silver or mercury species: thus we get rid of any anti-bacterial effects and modification from the genuine ionic composition of the aqueous medium. Calibration of this Pt pseudo-reference electrode was done by using aqueous solution of potassium ferrocyanide ($E_0$ = +360mV vs Norma Hydrogen Electrode - NHE)[28]: in this paper potentials ($V_{Pt}$) will be given versus our Pt pseudo-electrode ($V_{vs\ NHE} = V_{Pt} + 95mV$).

The electrical connection to AFM tip/cantilever system was made as following. The silicon platelet holding the cantilever was stuck on the JPK glass block holder by a two-component glue (purchased from JPK Instruments AG, Berlin, Germany), insoluble in water and culture medium. A polyester-imide insulated copper wire (purchased from Goodfellow, London, England), 100 microns in diameter, 11cm in length, was stripped at both ends (along few millimeters) with optical sand-paper to ensure electrical contacts with the potentiostat's cables (see below). This wire was then glued in the vicinity of the cantilever holder with the same product as that used for its fixation. One of the copper wire end was then approached to the Pt-covered silicon platelet and fixed on it by the deposit of a droplet of a conductive epoxy glue (Epoxy Technology USA) followed by curing at 110°C during 1 hour. Next a drop of nail polish was poured on the electrical contact and silicon cantilever holder and let dry at 40°C during one hour to electrically isolate the tip/cantilever complex from ionic solution. Thus the electric connection is made on the cantilever's side holding the tip. Electrodes were connected to a potentiostat (Modulab, Solarton analytical, AMETEK Advanced Measurement Technology, USA). For the current measurements an integration period of 150ms was chosen. Synchronization between time evolution of AFM data and electrical measurements from the potentiostat was done through short voltage pulse (10mV of amplitude) applied to the electrochemical cell.

## RESULTS

**Approach/retract curves**

A typical AFM height image of *Rhodo w.* is shown in figure 1.a where three associated bacteria are visible. The corresponding stiffness image is shown in figure 1.b. The typical height of those bacteria is in the range of 1µm. At the top-left corner of figure 1.a, a bump is visible. Its height is around 300nm as deduced from profile (figure 1.e) along white line in figure 1.a. This bump corresponds to a zone of high adhesion as seen in figure 1.c. Adhesion forces are there in the range of 5-6nN as seen in one particular retract curve (figure 1.d, orange curve) as measured at the red dot in figure 1.c. As explained in details in [10], this high adhesion zone corresponds to the presence of a polymeric substance likely secreted by the bacterium itself: this ensures its self-adhesion on the ITO/glass substrate. It must be noted that, in this particular medium (MM medium) with a high ionic strength (0.15M) no jump-to-contact is observed in the approach curve (figure 1.d, red curve), nor attractive long-distance interaction as expected when working in high salinity media as electrostatic forces are shielded[43]. Typical approach/retract curves from region of low adhesion zone (around 96% of the whole image) -see figure 1.c, blue dot- are plotted in figure 1.f: these curves are displayed for two adjacent pixels (full and dotted lines; these two pixels are distant of 31nm and the time interval between both is 12ms). Approach curves (in blue) reveal only two zones: the "long distance" zone (what we will call the "basis line", shortly mentioned as BL) where the force is constant at an accuracy of 15pN (see below) and the repulsive regime up to the set-up force at which retract of the cantilever begins. For the retract curve the repulsive regime is followed again by an almost constant force regime with superimposed oscillations: they are due to free oscillations of the cantilever as it abruptly goes off the adhesive contact. As seen in figures 1.d and 1.f, a shift along force axis between approach and retract curves is clearly evidenced. This is explained by hydrodynamic drag forces (viscous forces) due to the movement of the cantilever through the liquid medium. We indeed checked that this drag force, as characterized by the difference between extend and retract curves, (i) varies linearly with the pulling speed of the cantilever, (ii) has a constant value whatever the pixel's number, (iii) is proportional to the effective area of the cantilever (we checked it with various cantilevers as PPP-CONTPt, Nanosensors, Neuchatel, Switzerland and CSC37, MikroMasch, Wetzlar, Germany) and (iv) the measured values for this hydrodynamic drag force are in the same range of values as those reported in literature[44]. Due to that hydrodynamic effect, all retract curves are thus shifted from their corresponding approach curve, along force axis, by the *same* quantity for the constant conditions we imposed for AFM imaging. This fact combined to the lack of quasi-free oscillations in extend curves far away from the repulsive regime justifies why only the approach curves were analyzed.

**Definition of the so-called "basis-line force" (BL_force) signal.**

As already mentioned, the value of interaction force between the AFM tip and the substrate, during the approach phase, is constant for distances higher than these corresponding to the repulsive regime. Its mean value is called the BL_Force signal and will be noted as $F_{BL}$((see figure 1.b for a schematic).. Its exact determination is explained in details in Supplementary Information and figure SI01. The standard deviation for $F_{BL}$ was proved to be lower or equal than 15pN for all pixels of all AFM data presented in this paper and quantifies the accuracy of the so-called BL_Force measurement. The laser beam reflected by the AFM cantilever was aligned with the center of the four-quadrants photodiode at the beginning of each experimental session (corresponding to every studied set of bacteria) and thus $F_{BL}$ signal is equaled to zero at this time *t=0*. As shortly explained in introduction part, the variations of $F_{BL}$ is sensitive to variation of curvature of cantilever as caused by changes, for example, in electrical state of the cantilever and its ionic environment as it will be detailed below.

**Two classes of electrical status for AFM experiments.**

AFM experiments we will now discuss were performed under two different electrical conditions: The first one was done when no voltage is applied to the sample neither current measurement performed: we called this regime as "Open Circuit" (O.C.). The typical example we studied is that of consortium of three bacteria we already shortly described (figures 2.a-f). The second one was for AFM experiments where the voltage was kept constant and equal to 0mV versus the reference electrode and the intensity of current was measured by the ammeter/potentiostat simultaneously with AFM acquisition. One typical example is that reported for the case of a consortium of two bacteria as shown in figures 2.f-g. It must be emphasized that standard way of plotting AFM data by commercial softwares to get height images, rubs out any effect of possible variations of what we called BL_Force effects as data are shifted to a constant level of "BL_Force". We also verified that AFM data –in height, stiffness and adhesion modes- are independent of the electrical status we worked with, consequently proving that no perturbation in (i) AFM acquisition and imaging and (ii) in bacterial metabolism is engendered, at least at first order.

In order to check the electrochemical status in our experiments we performed cyclic voltammogram in a narrow window of potentials. One example is shown in figure SI02. It must be pointed out that such similar curves were measured whatever the point of observation (above or beside the bacteria) either with the AFM cantilever oscillating in QI mode or static at a constant distance (500µm) from the substrate. No difference was observed when digitization rate was changed. The main conclusion is that we do not observe any faradic current for this range of voltage: only a quasi-linear behavior is observed. As explained in reference [28] its slope (estimated from figure 3 to 1.5MΩ$^{-1}$) corresponds to an equivalent resistance between CE and REF electrodes. This resistance is due to two series connected resistances: one due to the contact resistance between the conductive electrodes and solution and the other to the own conduction of the aqueous medium. This last one can be estimated to 1.1kΩ (conductivity of culture medium equal to 14mS/cm). The first one is mainly due to polymeric substances constituting the bacterial membrane and extra-cellular substances partly involved in self-adhesion on the substrate.

**Results in the so-called "fast" regime.**

The first case we will present is that of a constant voltage (0mV) relatively to REF electrode. It is illustrated with the consortium of two bacteria (figures 2.g-h). As detailed above, $F_{BL}$, was measured during the AFM scanning of this bacterial consortium. Raw variations of $F_{BL}$, along the fast axis (horizontal in the shown AFM images) are plotted in figure 3 at four locations as indicated by the red lines in the simultaneously acquired stiffness image (see inserts in gray levels): the positions numbered 1 (4 respectively) in figure 3.a (3.d resp.) correspond to the bare substrate. The positions numbered 2 and 3 (see figures 3.b and 3.c.) are related to scan lines upon bacteria. The green and blue curves correspond to two adjacent (successive) fast scan lines. Out of the bacteria (figures 3.a and 3.d) these green and blue curves are superimposed at an accuracy of 15pN, the estimated error for $F_{BL}$. These experimental data are slightly dispersed along a linear fit of $F_{BL}$ ( dashed black lines) for the whole scan line. On the opposite, figures 3.b and 3.c reveal an important original feature: a lower $F_{BL}$ signal is measured for pixels upon the bacterial complex when compared to the linear approximation as calculated from both sides of the scan line out of bacteria. The mean value of that slope when averaged on all scan lines as acquired during the whole set of experiments on a bacterial consortium (see for instance figure SI03) is equal to ≈ -7±1pN/µm corresponding to a rate in time space of around -10±2pN/s. This is attributed to an un-compensated thermal drift as it will be seen with more details in the following. It must be noted that the signals of high intensity at the edges of the bacteria are related to a computing effect as explained with more details in supplementary information (see caption of figure SI01). To further explicit this new feature, the lower $F_{BL}$ signal upon bacteria, we define the $\delta F_{BL}$ signal, as calculated, at each pixel, by the difference between values of the $F_{BL}$ signal and of the corresponding linear fit ( black dashed line in figures 3), as taken at each pixel (see figure 3.b). That $\delta F_{BL}$ signal is plotted in figure 4. A net variation of the $\delta F_{BL}$ signal is clearly visible upon bacteria as it is shown in figures 4.a (three bacteria consortium) and 4.c. (two bacteria). For the complex of two bacteria as studied at 0mV relatively to REF electrode, the mean value of $\delta F_{BL}$ over the bacteria is equal to $\delta F_{BL}$ = -40±12pN (figure 3.a). Such a behavior was equally observed in a quite different electrical state. The three bacteria consortium was studied while no voltage was applied nor current measurement performed. However similar features in the $\delta F_{BL}$ signal were detected (figure 4.c). The amplitude of this signal is similar to that measured in case of constant voltage (0mV): $\delta F_{BL}$ = -50±15pN. Furthermore we checked that this $\delta F_{BL}$ signal is not directly linked to a spurious AFM effect where local variations of height would induce parasitic variations of $\delta F_{BL}$. We indeed plot in figure 4.b the $\delta F_{BL}$ signal corresponding to that in figure 4.a for only points higher than 100nm above the substrate's level: we can see that the thick exopolymeric substance with high adhesion properties (see figure 2.c) gives no $\delta F_{BL\ L}$ signal ($\delta F_{BL}$ ≈ 0) what is quite different from that detected over bacteria (~ -50pN). Same kind of observation was made with the case of two bacteria (see figure 4.d).

It is worth to note that this observation of a negative $\delta F_{BL}$ signal above the bacteria was observed whatever the AFM scanning conditions: this is detailed in supplementary figure SI03. Upon bacteria $\delta F_{BL}$ signal is almost constant even if heterogeneities are visible in figure 4: they will be addressed below. Typical time of variation of $\delta F_{BL}$ signal between zero level on substrate and that on bacteria is in the range of 30ms. This is shorter than the characteristic time of the second phenomenon we will discuss in next paragraph. Thus this first effect will be called "*fast effect*".

**Results in the so-called "slow" regime.**

In this sub-section the evolution of the raw BL_Force signal versus *time* is detailed. When $F_{BL}$ *(t)* is plotted for a complete sequence of successive AFM data acquisitions (figure SI04) a continuous linear drift is clearly observed whatever the scanning conditions. It is attributed to an un-compensated thermal drift. Its mean slope equals -8±2pN/s in case of constant voltage (two bacteria; see figure SI04.b) and is very close to the value obtained from study in figure 3. In another set of experiments where no electrical measurement was done (case of the consortium of three bacteria, fig. 2) this slope is equal to-1.5pN/s. In the following this thermal drift was systematically removed from the raw BL_Force signal: as for data in figure 3 (and SI03) we will now plot $\delta F_{BL}$ signals . In order to get rid of previously described fast effect, a smoothening of $\delta F_{BL}$ signal was systematically done: it corresponds to a temporal filter with a time constant of 150ms, equivalent to scanning time of twelve successive pixels. As mentioned in Material and Methods section, the analogic time constant for current measurements was set at a similar value (150ms).

For the case of AFM study at constant voltage (0mV), time variations of height (raw data), $\delta F_{BL}$ (smoothened as described above) and current intensity signals are plotted in figures 6.a-c. As height data are not temporally filtered the saw teeth corresponding to the AFM fast axis scan are clearly visible. The envelope of these height data gives the rough shape of bacteria end allows to situate the point of AFM scanning in the time diagram we will now describe in more details: as an example the numbered markers correspond to the four scanning lines studied in figures 3.a-d: on glass substrate before scanning bacteria, on the down bacterium then the top bacterium and at last the glass substrate after the AFM overfly of bacteria. We will now show that a supplementary signal can be measured with both $\delta F_{BL}$ signal and current intensity providing that the AFM scan speed is below a characteristics time. Indeed for AFM data (case of two bacteria) taken at a digitization rate of 64 pixels per line (figures 5.a and 5.b) we observe that both $\delta F_{BL}$ and intensity signals are constant, at the accuracy of our measurements, during the whole AFM scanning. This was observed independently of surface scan size, (5µm)² for image 5.a and (4µm)² for image 5.b. Surprisingly, when digitization rate reaches 128 pixels (figure 5.c, scan size: (4µm)²-) it is worth to note that both signals are no more constant. These two signals noticeably increase in a very similar way. We found a total variation of $\delta F_{BL}$ signal, $\Delta(\delta F_{BL})$ (as defined in figure 5.c), is in the order of 400pN while current intensity increases by roughly 20nA starting for a constant level of 172±2nA before bacteria scanning. This last value corresponds to current intensity as measured at 0V during voltage cycling (see voltammogram in figure SI02). This new effect will be called "slow" effect as it is clearly evidenced by applying a high-pass time filter, the cut-off time of which is about 150ms (integration time for current measurement and temporal smoothening for $\delta F_{BL}$ signal as well). The fast effect described above is thus not observable in the data presented in figure 5 as its characteristic time (30ms) is lower than the applied time filtering.

Remarkably we observed that this slow effect is observable by the lone measurement of $\delta F_{BL}$ signal. Indeed when working in the so-called open circuit (O.C.) configuration, identical observations (see figure SI05) to these detailed above can be made. The slow effect is again revealed for the only 128 pixels/line digitization rate (figure SI05.b) and not at 64pixels/line (figure SI05.a). This slow effect is characterized (figure SI05.b) by same type of variation of $\delta F_{BL}$ versus time within same range of magnitude, $\Delta(\delta F_{BL}) \approx 400pN$, when AFM tip overfly the bacterial consortium. When no time filtering is applied the fast effect is easily observed as already mentioned above (see figures 4.a and 4.b). In order to figure out in a better manner what is occurring during that AFM scan at a digitization rate of 128 pixels per line, the only condition when the slow effect is evidenced in our experimental conditions, we did a similar treatment as that shown in figure 3. Detailed comments are available in supplementary information (see also figure SI06).

Former described experiments were done with alive bacteria in their physiological medium. We also studied these *R. wratisl.* bacteria after they were starved to death: they were put in the electro-chemical cell filled with a pure NaCl solution (0.15M) - i.e. without any nutriment- during a minimum of 5 hours before AFM imaging. In these conditions neither slow nor fast effects on $\delta F_{BL}$ signal were observed whatever the digitization rate (figure SI07). It must be noted too that the related voltammogram, taken in identical conditions to these described in Material section, is quite similar to that shown for alive bacteria (figure SI02).

From this simultaneous observation of this so-called slow effect on both BL_Force signal and current intensity following remarks can be made:

(i) the observed signals in $F_{BL}$ (or $\delta F_{BL}$) are not due to artifacts due to signal processing and computing of AFM data as it is observed by two independent measurements, one based on direct measurement of electric current, the other by the detection of a mechanical effect (the flexion of the AFM cantilever);

(ii) as shown in figure 5.c, in the case of the slow effect, *variation* of $\delta F_{BL}$ signal is associated to the appearance of an over current when compared to the base current stemming from equivalent circuit of the electrochemical cell (as explained with more details in Material section; see also figure SI02). The $\delta F_{BL}$ signal is shown to be proportional to this over-current as clearly shown in figure 5.c. Thus the physical origin of *variations* of the $\delta F_{BL}$ signal is very likely due to the change of state of electrical charging of the cantilever electrode, generating a current detectable by the ammeter when the electric circuit constituted by ionic solution, conductive electrodes and wires is closed.

(iii) in the frame of this hypothesis it is possible to write that:

$$\Delta I \approx \frac{\Delta Q}{\Delta t} \approx \beta \cdot \frac{\Delta F_{BL}}{\Delta t} \qquad (1)$$

Where $\Delta I$ is the over current due to charging of $\Delta Q$ charges during time $\Delta t$, $\Delta(\delta F_{BL})$ the related change in $\delta F_{BL}$ signal. $\beta$ ratio can be estimated from experiments revealing the "slow" effect (figure 5.c). We will now show that from our hypothesis it is possible to justify that the fast effect cannot be detected by direct intensity measurements. The error bars for current and $\delta F_{BL}$ was indeed estimated to 2nA (fig. 5.c, orange curve) and 40pN (fig. 5.c, black curve) respectively. The characteristic time above which current variation is measurable was 150ms (see Materials section) : $\Delta t \approx 150 ms$. From these data $\beta$ ratio can be calculated. The so-called fast effect (effect N°1) is characterized by an amplitude of $\left|\partial F_{BL}^1\right| \approx 15\, pN$ (figure 3.b) for a typical time of variation of $\partial t_1 \approx 70 ms$ leading, by using equation (1), to an equivalent current of $\partial I_1 \cong 2nA$. This is the order of magnitude of error bar for the experimental determination of current intensity error (figure 5.c); thus the fast effect can be detected by our experimental set-up by intensity measurements.

We thus suggest that the BL_Force signal is related to the electric charging of the AFM tip/cantilever complex when contacting bacteria. We will develop this hypothesis and argue in this direction in the next section. Consequences on local measurements of electrophysiological properties of bacteria will then be presented.

## DISCUSSION

AFM is a powerful way to image in real space various substrates in various media with an sub-nanometric resolution. This method basically lays on the measurement of the small deflections of a microcantilever. That deflection is mainly due to two factors: the first one is the presence of an interaction (a force in the pico- or nano-newton ranges) between the apex of the tip and the underlying sample. The second comes from the variation of the difference of the surface stress between the two main surfaces of the cantilever. This second effect may come from specific adhesion on one side of the cantilever[37] or have a pure electrical origin as detailed below. Various kinds of interactions, such as electrostatic, van der Waals, magnetic, … with different distance dependencies, contribute to the total force between the probing tip and sample. The electrostatic force between two electrodes (the tip/cantilever system and sample) can be expressed[17] as

$$F_{El} = -\frac{1}{2}\frac{\partial C}{\partial z}V^2 \quad (2)$$

where $C$ is the capacitance of the system, and $V$ is the voltage difference. The capacitance depends on the geometry, distance, $z$, between the electrodes and dielectric properties of the medium. In gaseous conditions (in ultra-high-vacuum, nitrogen or air atmosphere) AFM scanning of various samples in electrical modes at a nanometric (or better) spatial resolution is currently reached in contact mode and non-contact mode as well[45,46]. In liquid, application of voltage between the tip and the sample induces redistribution of ions and water molecules so that a screening of electrodes mastered by diffusion processes (with a typical diffusion rate, $D$, of $\sim 4.10^{-9} m^2.s^{-1}$) occurs. The vicinity of the cantilever electrode in the electrolyte solution can be modelized in the frame of the Gouy-Chapman theory[47–49], as modified by Stern[50] to take into account the fact ions have a finite size: the metal electrode is surrounded by a so-called diffuse layer stemming from the attraction or repelling of mobile ions in solution by the metallic electrode according to its polarity and the opposite tendency due to thermal processes. This double layer (the diffuse layer and an inner layer –the Stern layer- of specifically adsorbed ions) screened the electrode potential on a characteristic distance named, $\lambda_D$, the Debye length. In case of the MM medium (ionic strength of 0.15M) we used in these experiments, $\lambda_D$ is small : $\lambda_D \sim 0.8 nm$. As a consequence the resulting electrostatic force as expressed by[15]

$$F_{El} = \frac{4\pi\sigma_t\sigma_S R\lambda_D}{\varepsilon_r\varepsilon_0}e^{-z/\lambda_D} \quad (3)$$

vanishes for very short distances and is only active in the strict vicinity of the repulsive regime. This explain that the value of the force we measured in the approach/retract curves is constant in the non-contact zone allowing us to define the so-called BL_Force signal for each pixel at an accuracy of 15pN. Such screening effect is currently used to increase AFM spatial resolution for samples imaged in ionic aqueous solutions[43] by adjusting the global interaction length. Thus direct electrostatic interaction is thus unable to explain the variations of the BL_Force signal we observed in both regimes.

The second origin for a modification of the flexion of the cantilever is, as said above, surface stress effects. Surface stress at the interface between the two phases, A and B, $\Gamma_{AB}$, is related to surface tension (surface free energy), $\gamma_{AB}$, by the Shuttleworth relation[31],

$$\Gamma_{AB} = \frac{d\gamma_{AB}}{d\varepsilon_\parallel} + \gamma_{AB} \quad (4)$$

with $\varepsilon_\parallel$ the elastic strain parallel to the interface. In case of a liquid/liquid interface, there is a strict equality between surface energy and surface stress[51] as Poisson ratio is equal to ½, meaning that the surface layer is incompressible. This is no longer the case when one of the phases is elastic. As shown in reference [51], compressibility of the interfacial region, through the Poisson ratio near the interface, determines the difference between surface stress and surface energy. However we will suppose in the following, as frequently done [52], that the following equation

$$\Gamma_{SL} \sim \gamma_{SL} \quad (5)$$

is valid in the present experiments.

One way to change surface stress is to vary the electrical potential of the solid or, more generally, the electrical state of the interface between a solid electrode and the surrounding ionic liquid (aqueous in the present case). By thermodynamical considerations[28,38] it is possible to derive the so-called *electrocapillary equation*[38]:

$$d\gamma = -\sigma.dV + \sum_i \Lambda_i.d\mu_i \quad (6)$$

where $\mu_i$ and $\Lambda_i$ are the chemical potential and the absolute surface excess (as defined in [28]) of species (labelled *i*) in solution. $\sigma$ is the excess surface charge density on the metallic side of the interface and *V* the electrode potential.

In case of no variation of these chemical potentials, equation (6) leads to the Lippmann's one:

$$d\gamma = -\sigma.dV \quad (7)$$

The property of varying surface stress by changes in electric state of solid/liquid interface was already drawn on with asymmetric AFM cantilevers [19,52,53] (a metallic layer on one side), standard commercial levers coated on both sides[17] and highly doped silicon cantilevers[53]. One important point revealed by papers by Umeda et al. [17,19] is that the cantilever behavior is dominated by surface-stress effects when frequency of the excitation signal is much lower than a characteristics frequency $f_{c\_ss}$[17]

In our case (ionic force of 0.15M and an AFM tip and bacterium heights around few microns,) $f_{c\_ss}$ is near of 35kHz. That value is thus far much higher than the typical frequency relevant for the experiments described in this paper: $f_V$ = 83Hz for the forced oscillations related to the approach/retract curves. Thus surface stress effects are dominant. Furthermore the fact that the magnitude of the surface stress only depends on the difference in the surface properties of both surfaces (tip side and backside) of the cantilever - and not from the distance between the cantilever and the counter electrode as with electrostatic force - well explains our observation of a tip-sample interaction force constant during the most of each approach curve. This feature further proves that the experimental results we described above are related to surface stress effect.

As mentioned in material and methods section the electrical connection with the cantilever is directly made on its side holding the AFM tip, the electro-active part of the cantilever. So an increase in stress of this surface due to a change of surface electrical properties of the AFM cantilever will be detected by photodiodes as a more attractive force and thus causes a decrease of the $\delta F_{BL}$ signal. Relationship between the (absolute) value of $\delta F_{BL}$ signal and the surface stress can be deduced from the Stoney's equation[54,55]. We chose a Young's modulus and a Poisson's ratio of 150GPa and 0.3 respectively, typical values for standard silicon cantilevers[37]. The length, width and thickness of the cantilever have been taken to 450 μm, 50μm and 2μm respectively. Surface stress and $\delta F_{BL}$ variations are then related through $\Delta\Gamma$ (in mN/m) = -1.5±0.5.10$^{-3}\Delta(\delta F_{BL}$ ) (in pN). Thus the module of variation of surface stress for the fast effect ($|\Delta(\delta F_{BL})| \approx 40 pN$) is in the range of 56μN/m.

As mentioned earlier we hypothesize that the variation of $\delta F_{BL}$ signal is related to an uptake of electric charges by AFM tip when contacting the bacteria. We will show now that this hypothesis is fully compatible in sign and magnitude with data currently available about electric charge surface density for standard bacteria. In the case of the experiments (figures 4.a and SI03.a-c) done at a constant potential, 0mV versus our REF electrode(i.e. $V_{vs\ NHE}$ = + 95mV), this potential is higher than the potential of zero charge, $V_{PZC}$. Raiteri et al.[52] indeed found negative values for $V_{PZC}$ for the AFM cantilevers, similar to ours, they used: between -500mV/NHE for a gold electrode in 0.1M KCl and slightly below 0V for Pt electrode in NaClO$_4$/HClO$_2$ solution. It means that[28], in experiments presented in this paper, there is thus a net positive surface charge on the cantilever. When referring to the fast effect (figures 4 and SI03) we observe a *decrease* in $\delta F_{BL}$ signal upon bacteria. According to preceding remarks, this corresponds to a *decrease* of the $\sigma.V_{/V_{PZC}}$ term in Lippmann equation (equ. (7)). As voltage may be considered constant, this corresponds to an uptake of *negative* charge by the AFM tip when upon the bacteria. From $\Delta(\delta F_{BL})$ variation (~ -40pN upon bacteria for the fast effect, see figures 3 and 4) a rough estimation of the uptake charge can be proposed. We suppose that the charge uptake occurs when the two double layers are in contact, it means, mainly during the repulsive phase of the approach/retract curves (figure 1). This lasts $\tau \approx$ 0.5ms as deduced from extend/retract curves. During that time lapse, ions are supposed to stem from a interfacial area corresponding to a surface limited by diffusion, $\sim D.\tau \cong 2\mu m^2$, and to charge the surface of the cantilever (area $L.\ell$, with $L = 450\mu m$ and $\ell = 50\mu m$). By using Stoney[54,55] and Lippmann equations successively we get a rough value of $\sigma_{bact} \sim -0.5\ C/m^2$ for bacterial surface density. For this surface density determination we supposed the electric potential is in the order of magnitude of the standard electrode potential in aqueous solution for platinum at 25°C, ~1V versus NHE [56]. The value for surface charge we obtained from $\delta F_{BL}$ signal, $\sigma_{bact} \sim -0.5\ C/m^2$ does satisfactorily agree with values from literature for Gram positive bacteria at neutral pH: for instance Bulard et al.[57] found around $\sigma_{bact} \sim -0.4\ C/m^2$ for *Lactococcus lactis* in NaCl at 1.5mM ; Poortinga et al[2] measured for several Gram-positive bacteria, including *Rhodococci*, surface charges ranging from $-0.2\ C/m^2$ to $-0.5\ C/m^2$ in 1mM KNO$_3$. It must be emphasized that our estimation of surface charge was done *in-situ* with alive bacteria in their physiological medium (150mM). We put the stress on the fact that this $\sigma_{bact} \sim -0.5\ C/m^2$ value we assessed does not depend on AFM surface scan size, number of pixels per line nor electric type of measurement (constant potential or open circuit). As visible in figures 4 and SI03, the distribution of the charges over the bacteria is rather uniform. Some surface heterogeneity is however visible: this feature will be detailed below. It must be emphasized that during this fast effect the $\delta F_{BL}$ signal and, in the frame of our hypothesis, the charge of cantilever, reaches a constant value as soon as the AFM tip is over the bacteria and then keeps a constant charge all over the surface of the bacteria. This observation may explain why no extra current will flow through the ammeter as seen in figures 5.a-b. On the opposite, during the so-called slow effect, a continuous increase of $\delta F_{BL}$ signal is observed corresponding to a total net change of approximately $|\Delta(\delta F_{BL})| \approx 400pN$ (figures 5.c and SI05.b), ten times higher than that detected for the fast effect. From that we can estimate the variation of the total charge $|\Delta Q| \approx 1.35\ 10^{-11}C$. From that it could tentatively be given an estimation for related current intensity flowing through system and ammeter. Indeed similarly to arguments invoked above for fast effect this charge increase might be considered to occur during an equivalent contact time between both double layers of $\tau \approx$ 0.5ms. This would lead to a net intensity current of roughly 30nA, a value of the same order of magnitude as that observed experimentally ~20nA (figure 5.c). These remarks contribute to strengthen the hypothesis we made of a uptaking of electric charge thanks to intermittent contact between both electric double layers, one from the underneath substrate and the other the AFM tip/cantilever complex and its indirect measurement thanks to an electro-mechanical effect based on the variation of flexion of the cantilever through electrocapillary effect.

We propose the following mechanism to tentatively explain how the processes of uptaking of electric charges by AFM tip when approaching the sample's surface. From studies made with electrode plunging and emerging from an aqueous electrolyte[27,29,58] it was shown that the electrochemical double layer can be retained intact on an electrode surface as the electrode is emersed[25] to air or vacuum as well. It was indeed proved that both the charge on the electrode and the potential across the emerging double layer remain fixed as this electrode is removed[27]. The model [25] that was proposed is based on the hypothesis of the "unzipping" of the EDL during electrode's emersion takes place just outside the outer Helmholz plane as defined in classical theory of EDL[28,29]. Furthermore it was detailed that immersion/emersion cycles of the electrode can be repeated many times without perturbation at any potential not dominated by Faradaic currents, on either side of the p.z.c. [27].

Let us come back to the so-called slow effect. We will now suggest a mechanism, related to bacterial metabolism, as a hypothetic explanation for the only detection of the slow effect in case of high digitalization rate (figures 5.c and SI05.b). It is well known that ion channels are widespread in prokaryotic membranes[59,60]. Some of these ionic channels can present mechanosensitive properties[61]. Numerous studies based on electrophysiological experiments have indeed shown that different types of mechanosensitive channels are present through the cytoplasmic membrane of bacteria. These channels (MscL, MscS, and MscM in Escherichia coli cells[62], MscCG in *Corynebacterium glutamicum* [5], *Vibrio cholerae*[6] etc…) can release intracellular molecules and ions to reduce osmotic pressure when the cells are challenged with osmotic stresses. Typical channel surface densities is typically in the range of 10 channels per micrometer-square[5,6]. Furthermore, depending on the activated channel, time constant of the opening of mechanosensitive channels is in the range few 10-100ms[5]. From that we can deduce that this effect might be detected in our experiment if the following condition is satisfied: the duration for electric contact between tip and substrate has to be at least in the range of 10-100ms (time of opening for one ionic channel) for a scanned surface of 0.1µm² (minimum surface to statistically overfly a ionic channel). As already mentioned the electric contact between tip and substrate occurs when both double layers are in interaction thus in the repulsive phase. This interaction time is equal to ~0.5ms per pixel. From that we deduce that the time spent by the tip in electric contact is ~15ms in case of digitization rate of 64pixels and ~50ms in case of 128pixels for an investigated bacterial surface area of 0.1µm² (AFM images of (4µm)²). Our experiments showed that there is a time threshold between these two limiting experimental cases (64p and 28p) we can place at few tens of milliseconds. This threshold is compatible with data from patch-clamp measurements that measured an opening time of ionic channels around 10-100ms as mentioned above. We are thus confident in the fact that the slow effect could be related to the detection of ionic current stemming from opening of channels at the bacterial membrane.

As seen in figure 6 heterogeneities in surface charge distribution are present. The stiffness image related to the height image in figure 1.a is plotted in image 6.a: zones of bacteria with a slightly lower stiffness are visible. We will focus our attention on the circular zone situated on the right bacterium as shown by the white or black circle in figure 6.a (stiffness data) or 6.b ($\delta F_{BL}$ signal). This heterogeneous zone is better evidenced on height profiles (figures 6.c and 6.d, green curves), stiffness curves (figure 6.c, black curves) and $\delta F_{BL}$ profile (blue lines in figure 6.d). Two profiles right across the protuberance (full line in fig. 6.a) and slightly above it (dashed line in fig. 6.a) are plotted in figures 6.c-d: the thick (thin respectively) lines are for section across (resp. above) the protuberance. This one protuberance is located between the two dotted-dashed black lines (figures 6.c-d). It is evidenced that the protuberance corresponds to a zone with a lower stiffness and a slight increase in $\delta F_{BL}$. This corresponds, according to former remarks, to a local charge less negative than in the main part of bacteria. This might be interpreted as a localized lack of teichoic acid leading to a decrease of electrical negative charge in relation with a lower stiffness of bacterial membrane[63]. This explanation can be advanced as we verified (see "results" section) that roughness and electro-mechanical effects are uncorrelated. The stress has to be put on the fact that these last observations are valid whatever the AFM digitization rate. Thus these measurements might reveal that the study of $\delta F_{BL}$ lead to important information about permanent state of charge of bacterial membrane. Furthermore this proves that BL_Force signal can provide information on local charge distribution at a spatial resolution better than few tens of nanometers.

It is worth to note that the experimental set-up was chosen in order to be able to correlate evolution of $\delta F_{BL}$ signal with that obtained from a direct measurement of electric current in well controlled electrochemical conditions. For that purpose a conductive substrate (ITO on glass) was used. We thus observed that $\delta F_{BL}$ variation in the, what we called, open-circuit condition was similar to that we measured with the cantilever in a well characterized electric state (at a null potential versus our reference). This shows that the electrical state spontaneously reached by the cantilever before AFM experiments (self-potential as often called) corresponds to a potential near of that of the experiment with direct electric measurement (0mV versus the reference electrode we used ). We thus proved that local measurement of surface charge - and their variation in time (due to the opening of ionic channels as hypothesized here)- does not need the use of ammeter as we showed they can be realized by the only measurement of the $\delta F_{BL}$ signal. In that last case a conductive substrate is not needed.

From preceding remarks, it can be anticipated too that disjoined polarization of both cantilever and a conductive substrate –via a bi-potentiostat- is likely feasible : interesting studies about effects of polarization on physiology of bacteria are thus possible.

In the experiments presented here, the applied potential (when electric circuitry was used) and the self-potential (in the opposite case) was higher than the potential of zero charge. Same kind of results can be obtained for potentials lower than $V_{PZC}$ : indeed in the experiment described in this paper, according to Lippmann's equation (equ. (6)) a decrease of $\delta F_{BL}$ is related to an uptake of negative charge whatever the relative potential of working potential versus $V_{PZC}$ provided that the resulting potential due to the charging does not change of relative position versus $V_{PZC}$ (as the electrocapillary curve is maximum at $V_{PZC}$).

It must be mentioned that typical times for extend and retract ramps of the cantilever have to be chosen in a narrow window: larger than that related to reach the charging equilibrium between the two facing double layers (estimated in the order of ≈10μs) and lower than those related to relaxation processes. As mentioned in paper of Raiteri et al.[52] where kinetic measurements were done, these later processes may be related to (i) a first effect with an intermediate time constant in the order of few tens of milliseconds and hypothetically attributed to a residual electrochemical effect and (ii) a slower component (around few tenths of seconds) tentatively related to a "diffusion controlled process". As mentioned above, in our experiments, a new charging equilibrium is reached every ~15ms at every ramping down of the cantilever therefore minimizing the role of these relaxation processes with longer time scales. Further studies in order to study in detail these time effects are under work.

## CONCLUSION

We reported an *in-vivo* electromechanical AFM study of charge distribution on the cell wall of Gram+ *Rhodococcus wratislaviensis* bacteria, naturally adherent to a glass substrate, in physiological conditions. The new method presented in this paper relies on a detailed study of AFM approach/retract curves giving the variation of the interaction force versus distance between tip and sample. In addition to classical height and mechanical (as stiffness) data, mapping of local electrical properties, as bacterial surface charge, was proved to be feasible at a spatial resolution better than few tens of nanometers. This was done by studying the constant level of the cantilever's bending far away (>10nm) from the contact zone between the AFM tip and the sample during approach of the tip to the sample, the so-called BL_Force signal. The resulting deflection of cantilever, due to surface stress variations, is coming, as in classical electrocapillary experiments, from variation of its surface charge density. The electrical charging is supposed to occur during the contact of the two electrical double layers the typical thickness of which is below 1nm as these AFM experiments are done in a high ionic strength of liquid media (0.15M). Estimation of electrical surface charge was done and proved to be compatible with results from standard macroscopic electrophoretic mobility measurements. Furthermore, an additional electrical signal detected by both the deflection of the AFM cantilever and simultaneous direct current measurements was detected at low scanning rates. It was tentatively attributed to the detection of current stemming from ionic channels the opening of which might be triggered by local mechanical overpressure generated by AFM contact. More work is now needed in order to better know the mechanisms for local charging of the cantilever.

This method offers an important improvement in local electrical and electrochemical measurements at the solid/liquid interface particularly in high-molarity electrolytes when compared to technics focused on the direct use of electrostatic force. The experimental results presented in this paper tends indeed to prove that the careful study of the BL_Force signal is an elegant way of performing patch-clamp-like experiments on alive bacteria in their physiological medium without the need of indirect method as the preparation of giant spheroplasts by lysozyme digestion of the native bacteria. The counterpart is the temporal restriction to variations of electromechanical signal slower than few tens of milliseconds. However it could be overcome soon as further studies to progress in this direction are under work. The method we detailed here thus opens a new way to directly investigate "in vivo" biological electrical surface processes involved in numerous practical and fundamental problems as bacterial adhesion, biofilm formation, microbial fuel cell, etc.

## ACKNOWLEDGMENTS

This work was supported by Agence Nationale de la Recherche, Paris, France within the framework of ECOTECH program "BIOPHY". The authors are grateful to IFP Energies Nouvelles, Rueil-Malmaison, France (Dr. Françoise

Fayolle-Guichard and Yves Benoit) for the free disposal of *Rhodococcus wratislaviensis*, IFP 2016 strain, through Dr. Fabienne Battaglia, Marie-Christine Dictor, Jean-Christophe Gourry and Caroline Michel, BRGM, Orléans, France. This work was launched at Geosciences Montpellier laboratory thanks to support of Dr. S. Lallemand, J.-L. Bodinier and P. Pezard. S.D. acknowledges support from Région Languedoc-Roussillon, France.

FIGURE CAPTIONS

Figure 1:
Schematics of experimental set-up: figure 1.a in case of measurements of both $\delta F_{BL}$ signal and direct electric current; figure 1.b in case of the lone measurement of $\delta F_{BL}$ signal.

Figure 2:
AFM images (128 pixels x 128 pixels), height data in figure 2.a, stiffness in Fig. 2.b and adhesion in figure 2.c , of a consortium of two Gram+ *Rhodococcus wratislaviensis* bacteria in their physiological medium. The height profile along white line in fig. 2.a is plotted in figure 2.e: the bump at the left side is due to an EPS layer ensuring the adhesion of the bacterium on the substrate. Approach/retract curves at the EPS zone (red point in image 2.c) and upon the bacteria (blue point in Fig. 2.c) are respectively shown in figures 2.d and 2.f. In this last case curves are shown for two adjacent pixels (full lines or dotted lines). In figures 2.g-h another consortium of two bacteria is depicted in height (figure 2.g) and in stiffness (figure 2.h) in (128 pixels x 128 pixels) images. The scale bar represents 1µm.

Figure 3:
Spatial variations of the raw BL_Force ($F_{BL}$) along two *successive* horizontal scan lines at four different positions over the two-bacteria consortium as indexed in inserts. The first acquired line is plotted in blue, the second in green. The AFM data corresponds to the image shown in inserts and acquisition conditions are: scanned area (4µm)²; digitization rate: (64pixels)². The black dashed lines are the best linear fits of profiles as determined in the portions without bacteria (left and right sides). The definition of $\delta F_{BL}$ is schematized in figure 3.b.

Figure 4:
Images calculated from the $\delta F_{BL}$ signal are plotted in figures 4.a and 4.c for the three bacteria and two bacteria consortia respectively. . The scale bar represents 1µm. The digitization rate was (128pixels)².
Figures 4.b and 4.d are portions of images 4.a and 4.c, respectively, corresponding to height data higher than 100nm above the substrate's level.

Figure 5:
Variations of $\delta F_{BL}$ signal(black lines), current intensity (orange curves) and height (same color code as in figures SI04.a-b) signals versus time for three AFM scanning conditions: fig. 5.a: scanned area (5µm)²; digitization rate: (64pixels)²; fig. 5.b: scanned area (4µm)²; digitization rate: (64pixels)²; fig. 5.c: scanned area (4µm)²; digitization rate: (128pixels)². Potential was kept constant and equal to 0mV versus the pseudo-reference electrode. Spatial variations of $\delta F_{BL}$ along horizontal scan lines starting at times noted by numbered markers in figures 5.b and 5.c are shown in figures 3 and SI06 respectively.

Figure 6:
AFM stiffness and $\delta F_{BL}$ images of bacterial consortium of figure 2.a are plotted in figures 6.a (see also Fig. 2.b) and 6.b (see also Fig. 4.b) respectively. The zone circled by a white line in stiffness image –and by a black one for $\delta F_{BL}$ data- is studied in more details in profiles plotted in figures 6.c-d.: the profiles in thick line are taken along the yellow full line in image 6.a; it characterizes the heterogeneous white-circled zone visible in Fig. 6.a. This heterogeneous zone is located by the two dot-dash lines in figures 6.c-d. In figures 6.c-d, height profiles are plotted in green, stiffness in black and $\delta F_{BL}$ in blue lines. The profiles in thin lines (Fig. 6.c-d) are plotted along the dashed yellow line in figure 6.a and are characteristic of points out of the white-circled zone. It is thus shown that surface charge heterogeneity is linked with bacterium morphology.

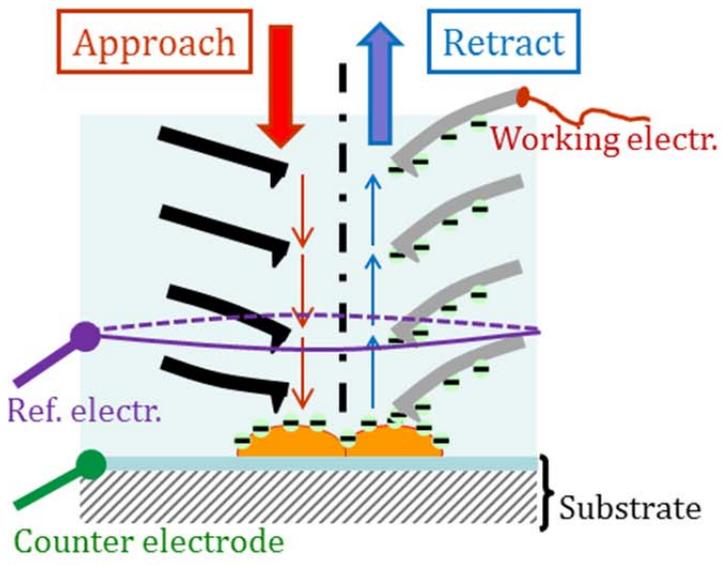 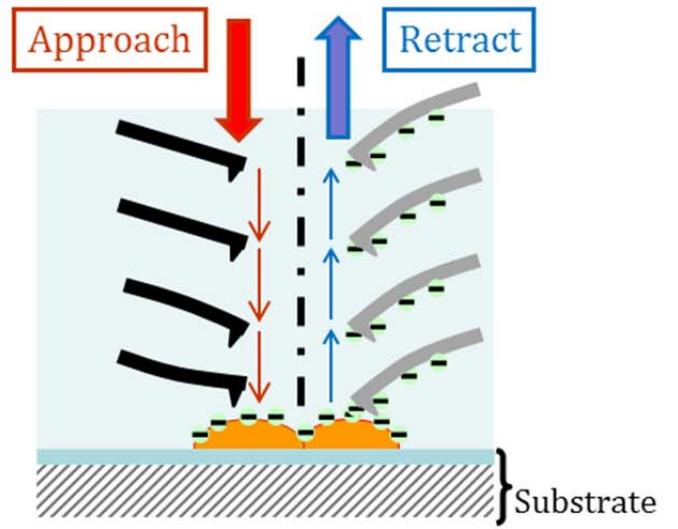

Figure 1.a　　　　　　　　　　Figure 1.b

Figure 2.a

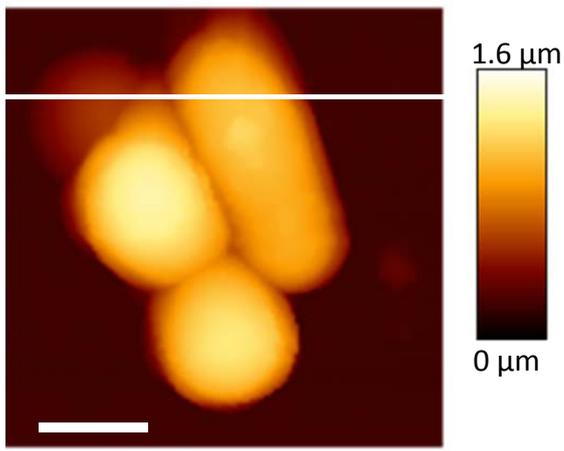

Figure 2.b

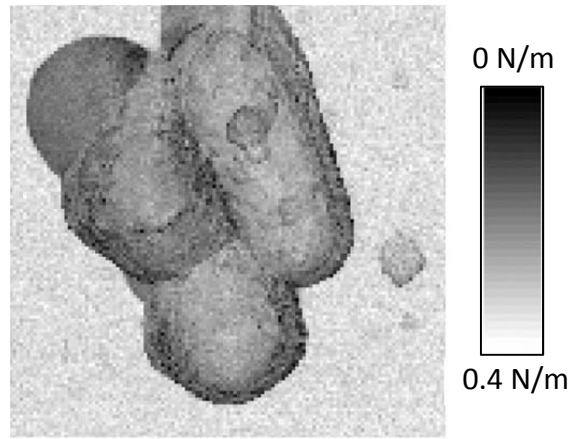

Figure 2.c

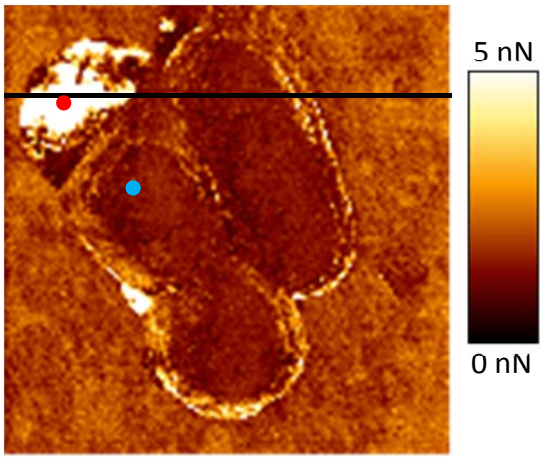

Figure 2.d

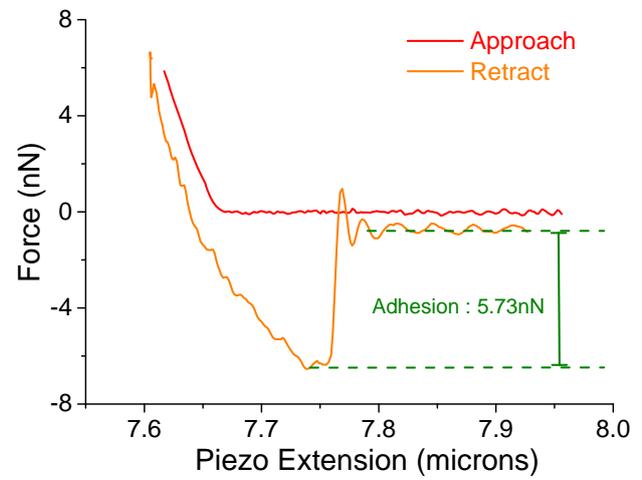

Figure 2.e

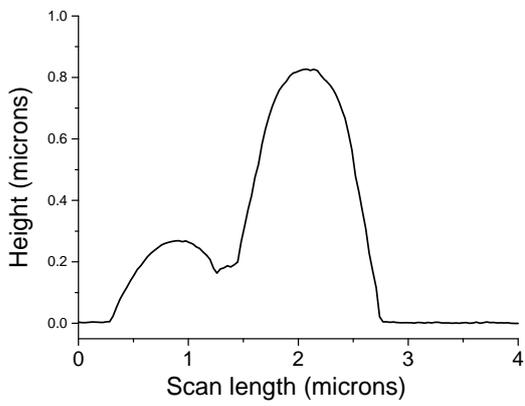

Figure 2.f

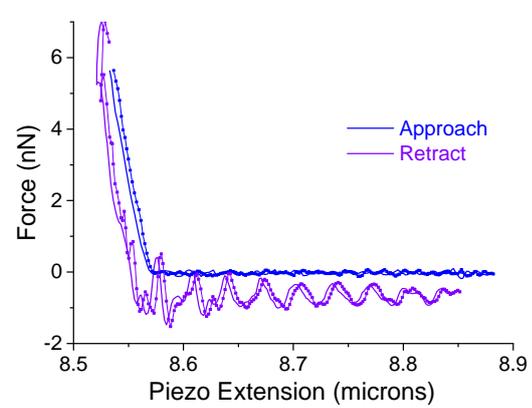

Figure 2.g

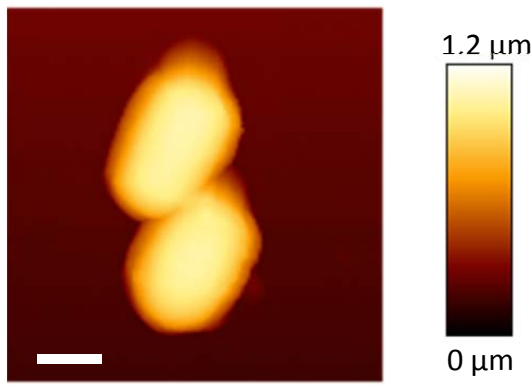

Figure 2.h

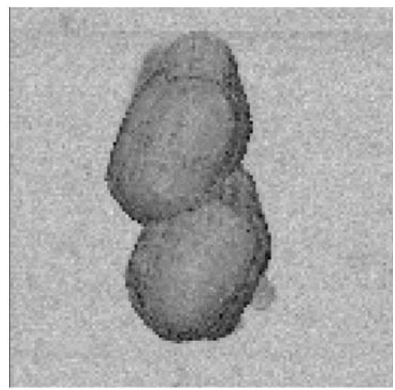

Figure 3.a
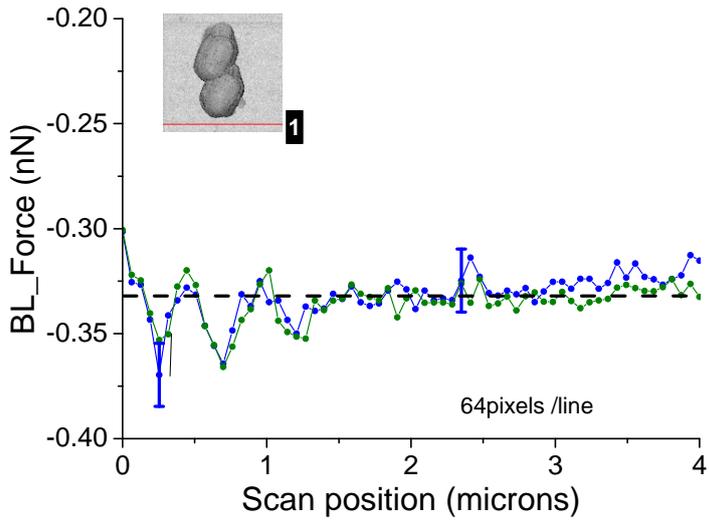

Figure 3.b
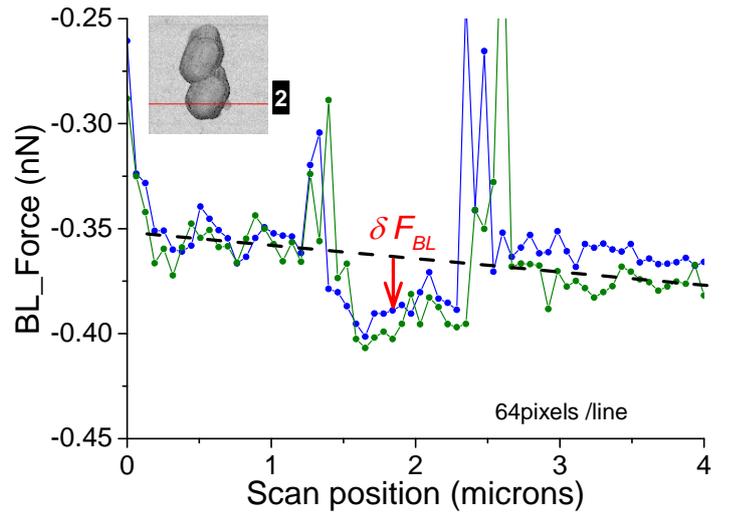

Figure 3.c
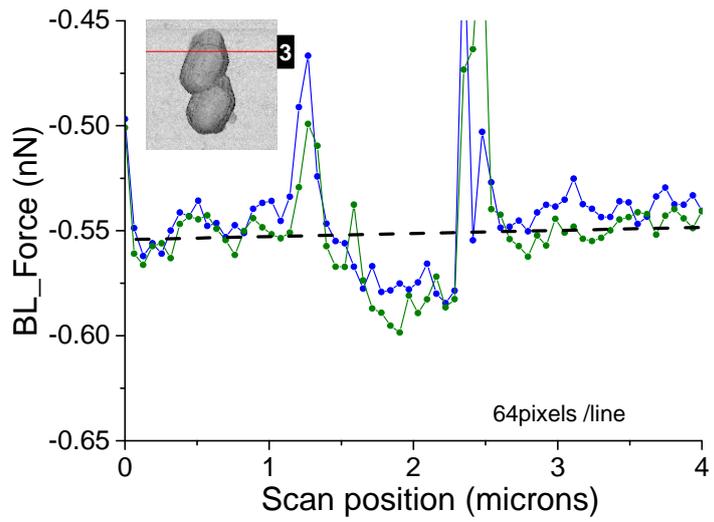

Figure 3.d
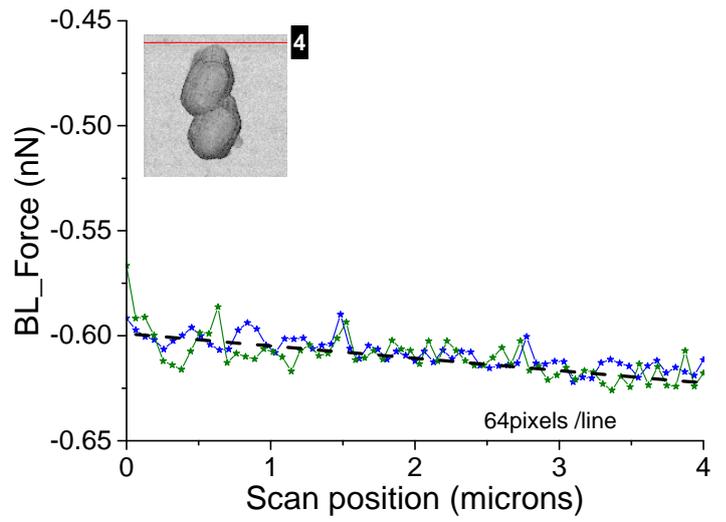

Figure 4.a

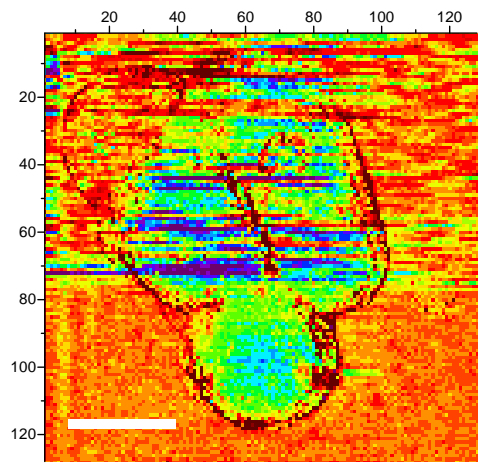

Figure 4.b

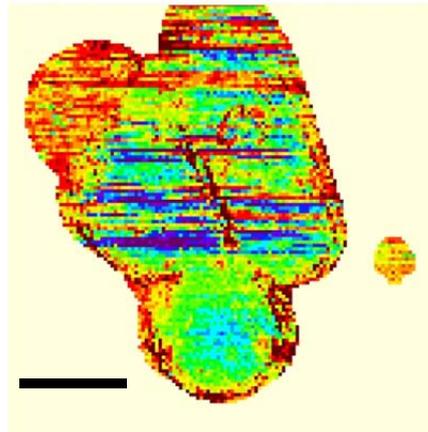

Figure 4.c

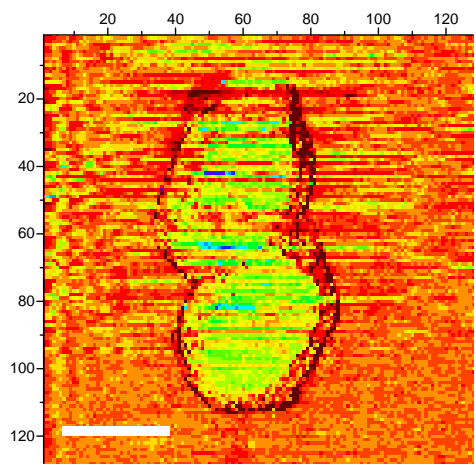

Figure 4.d

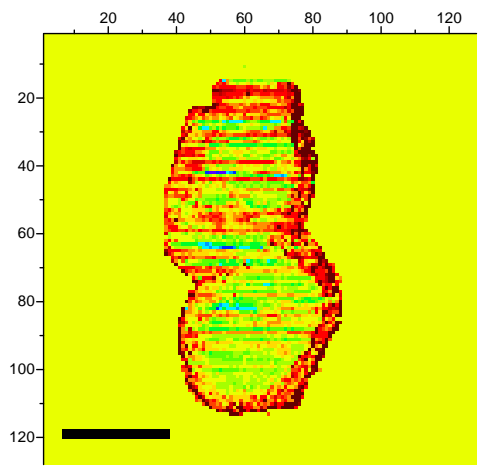

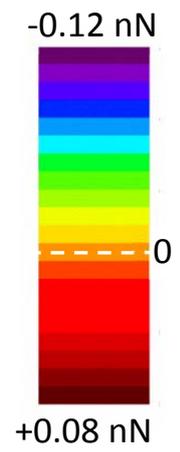

Figure 5.a

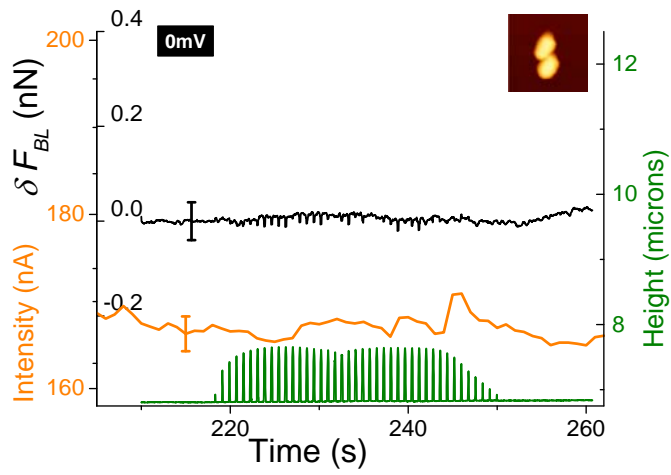

Figure 5.b

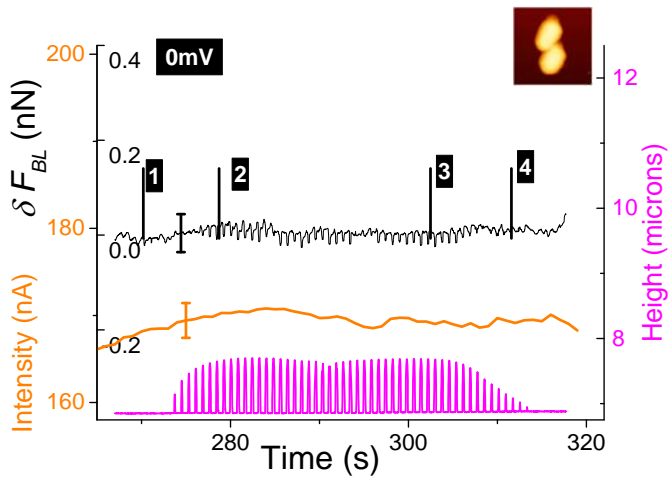

Figure 5.c

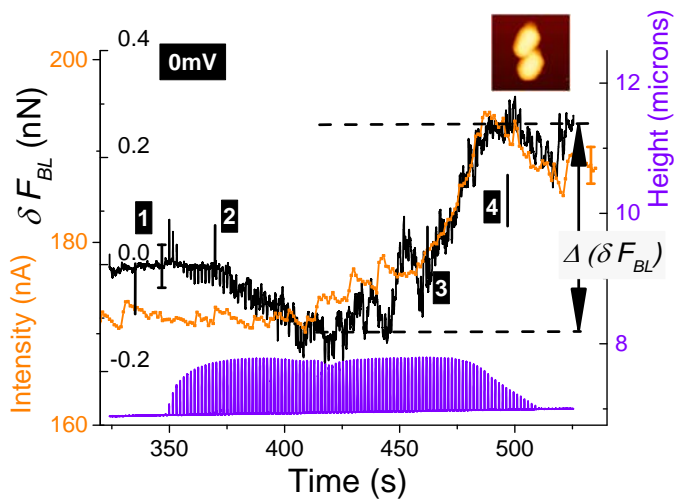

Figure 6.a

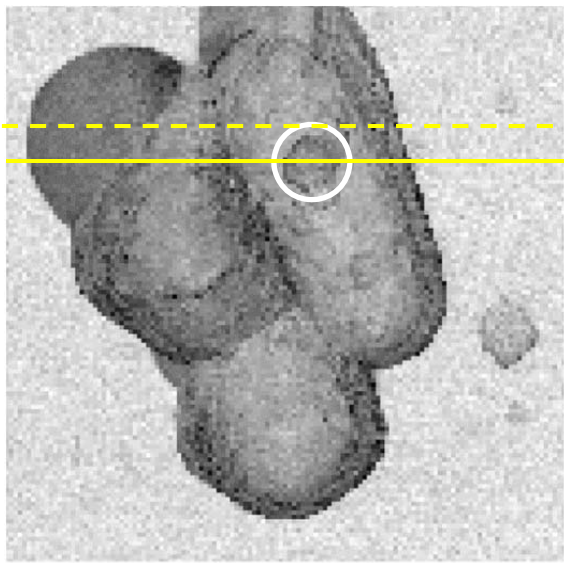

Figure 6.b

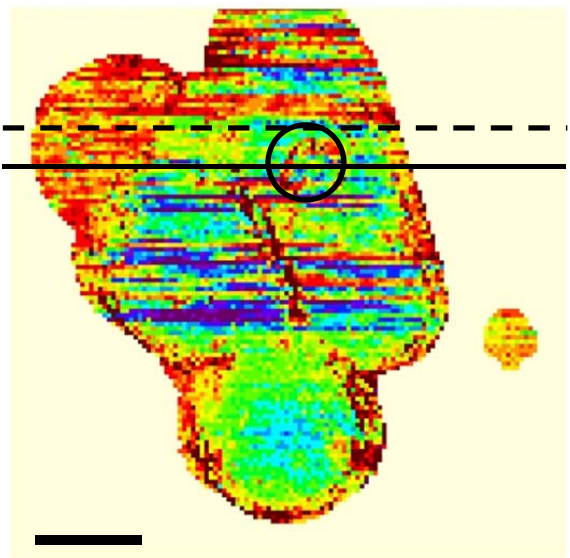

Figure 6.c

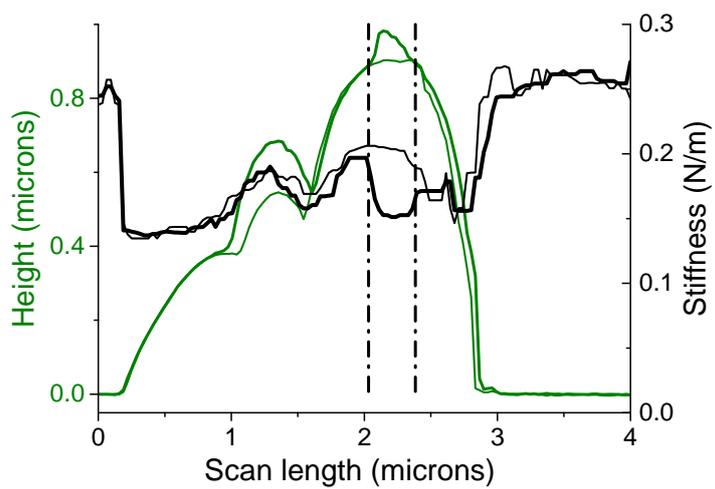

Figure 6.d

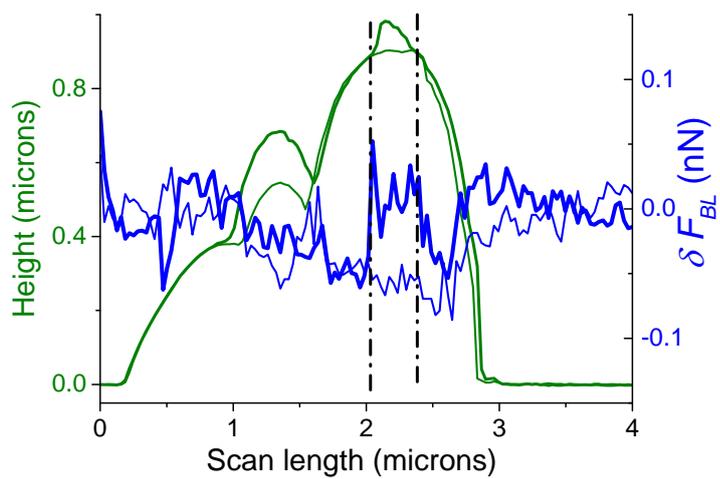

# Supplementary Information

Detailed explanation for calculation of the BL_force signal denoted by $F_{BL}$

The force signal in approach (extend) mode is linearly fitted by two segments (see figure SI02): the first fit is obtained along a piezo-extension of 10nm at point of maximum force (~6nN). The second comes from the linear fit along a segment, 200nm in length, starting from the most distant point in the extend curve. The value, along the "piezo-extension" axis, of the intersection between these two segments is the lower bound for the calculation of the average value of $F_{BL}$, the BL_Force signal. The upper bound is the most distant point in the extend curve. It must be noted that the absolute value of the slope for the best linear fit of the constant portion of F(d) curve is lower than 30pN/micron whatever the pixel.

Study of intermixing of "fast" and "slow" effects

Figure SI06 shows variations of raw BL_Force signal ($F_{BL}$) along the fast axis for the *same* four locations as in figure 2. No temporal filter is applied and fast variations of the raw signal can be detected as for figure 2. Two successive fast-scan lines are shown: the first in magenta, the second in red. Equivalent results to those for the low digitization rate are clearly visible on the first two images (figures SI06.a-b) i.e. before AFM tip reaches the first third of the bacterial complex. The $F_{BL}$ signal reveals as for 64 pixels case a quasi-linear variation of signal (figure SI06a) with a slope of roughly -15pN/s as in figure 2. We can again attribute it as for the case of 64 pixels digitization rate to an un-compensated thermal drift. The so-called "fast" regime is still present in $F_{BL}(t)$ curves as $\delta F_{BL}$, as defined in case of a digitization rate of 64 pixels, still has a negative value upon the bacteria and comes back to zero away from them. This explains again the observed contrast in reconstituted images based on this $\delta F_{BL}$ signal (figure 3.c). The new and important point is that when AFM tip starts to scan over the bacterial aggregate we observe that raw BL_Force signal, $F_{BL}$, is strongly perturbed: the slope of $F_{BL}(t)$ curve becomes positive leading to a correlative increase of $F_{BL}$ versus time as observed in figure 4.c. This effect ("slow" one) occurs at a time scale higher than a threshold we can roughly estimate to be equal to the duration of one scan line with 64 pixels, ~800ms.

SUPPLEMENTARY FIGURE CAPTIONS

Figure SI01:
This figure, plotting the interaction force between the AFM apex and the sample versus the elongation of Z-piezo illustrates how the BL_Force signal, $F_{BL}$, is calculated. The force vs elongation curve in extend mode is fitted by two straight segments: the first one (dashed lines) is obtained from a linear fit of the repulsive region starting from point of maximum force (~6nN) and over a 10nm range. The second comes from a linear fit, parallel to the piezo extension axis, along a segment, 200nm in length, starting from the most distant point in the extend curve. The $F_{BL}$ signal is the mean value of the interaction force calculated along the constant part of the F(d) curve between the two following bounds : the intersection between the two former linear fits and the most distant point in the extend curve (domain of calculation is indicated by horizontal arrows in Fig. SI01). The red curve is typical to almost every data point of the AFM image except those at the edges of bacteria. In this case a typical *F(d)* plot is drawn with the blue line. The bump visible at the foot of the repulsive domain is due to increase lateral interaction between the tip and the edge of bacteria; it explains the presence of spurious increase of BL_Force signal ($F_{BL}$), as calculated by the automatic procedure, along edges of bacteria as visible in figures 2 and SI06.

Figure SI02:
Variations of current versus voltage as measured during AFM acquisition in a small zone (10nm square) of the two bacteria consortium of figures 2.g-h. Same kind of voltammogram was obtained for a very large panel for conditions of measurement: see text for more details.

Figure SI03:
Images calculated from the $\delta F_{BL}$ signal are plotted in figures SI03.a-c and SI03.d-e for the two bacteria (case of V = 0mV)and three bacteria consortia (case of no electrical connexion) respectively. The scale bar represents 1μm. These images were taken under following conditions:
fig. SI03.a: (5μm/64pixels)²; fig. SI03.b and SI03.d: (4μm/64pixels)²; fig. SI03.c and SI03.e: (4μm/128pixels)².

Figure SI04:
In figure SI04.a is described the temporal succession of images before the AFM scanning of image in figure 2.g. $1^{st}$: yellow line; scan size: 10μm. $2^{nd}$: blue line; scan size: 20μm. $3^{rd}$ : red line; scan size: 10μm. $4^{th}$ : green line; scan size: 5μm. $5^{th}$: magenta line; scan size: 4μm. Then a last image (see figure 2.g) was taken. All these images except that plotted in figure 2.g were scanned at a digitization rate of (64pixels)². Figure SI04.b: time variation of current intensity (orange line), raw BL_Force ($F_{BL}$) signal (the color curves near the black dashed line) and AFM height data (the upper colored curves) for successive AFM images according to the sequence described in figure SI04.a with the same color code. AFM data in violet corresponds to figure 2.g.

Figure SI05:
Variations of $\delta F_{BL}$ signal (black lines) and height signals versus time for two AFM (4μm)² images of the bacterial consortium shown in figure .a-c1 with different digitization rates: fig. SI05.a: (64pixels)²; fig. SI05.b: (128pixels)². In this case electrodes were not connected to the potentiostat (what we called the open circuit –O.C.- conditions).

Figure SI06:
Spatial variations of the raw BL_Force ($F_{BL}$) along two *successive* horizontal scan lines at four different positions over the two-bacteria consortium as indexed in inserts. The first acquired line is plotted in magenta, the second in red. The AFM data corresponds to the image shown in inserts and acquisition conditions are: scanned area (4μm)²; digitization rate: (128pixels)². The black dashed lines are the best linear fits of red profiles as determined in the portions without bacteria (left and right sides).

Figure SI07:
Variations of $\delta F_{BL}$ signal (black lines) and height (blue lines) signals (figure SI07.c-d) versus time related to the two AFM images (figure SI07.a-b) of non-living *Rhodococcus wratislaviensis* bacteria: they were studied by AFM five hours after the AFM electro-chemical cell has been filled with a pure NaCl solution (0.15M), i.e. without any nutriment. Scan size: (7.9µm)². Digitization rates: fig. SI07.a and SI07.c: (64pixels)²; fig. SI07.b and SI07.d: (128pixels)².

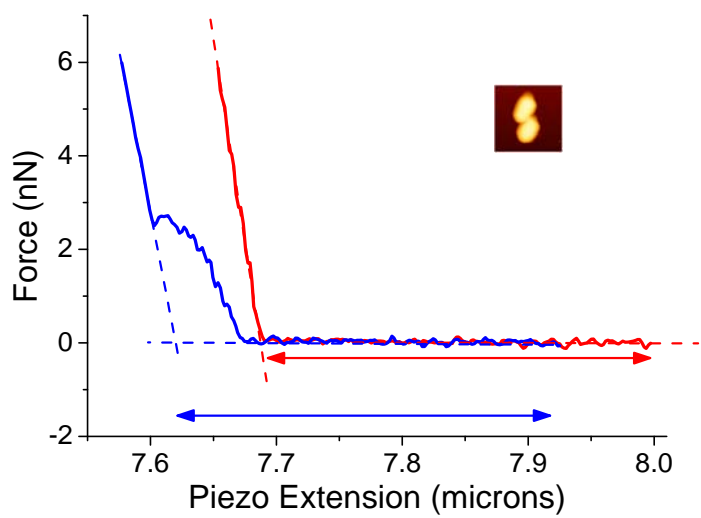

Figure SI02

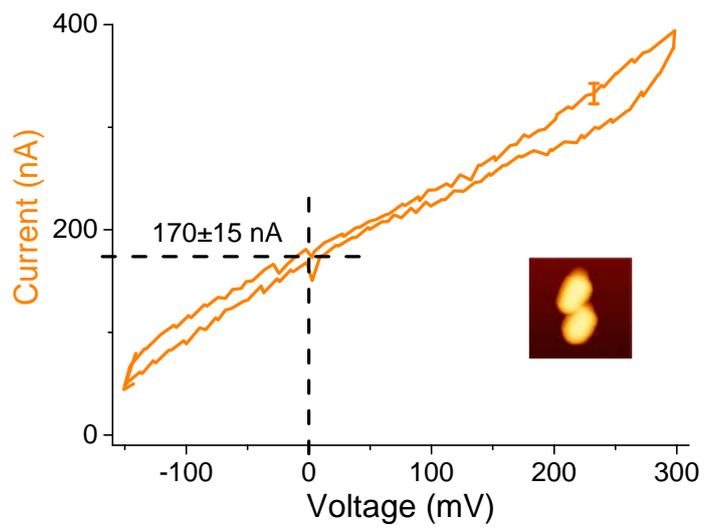

Figure SI03.a 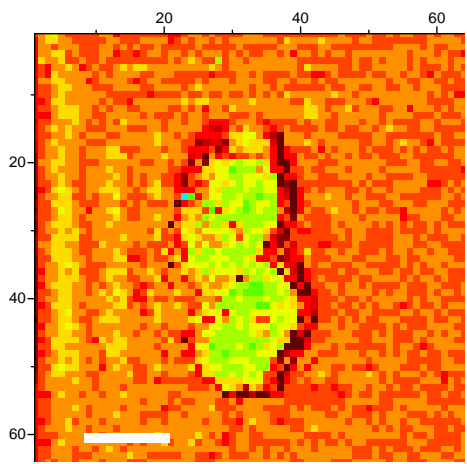 Figure SI03.b 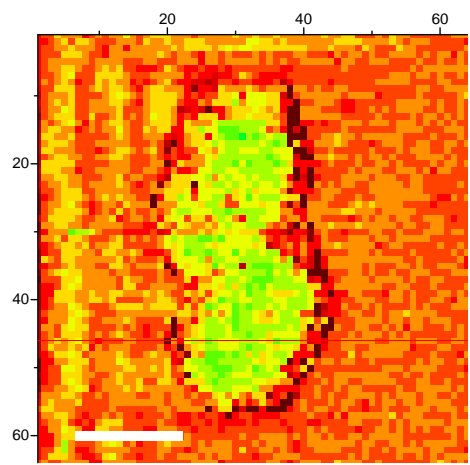 Figure SI03.c 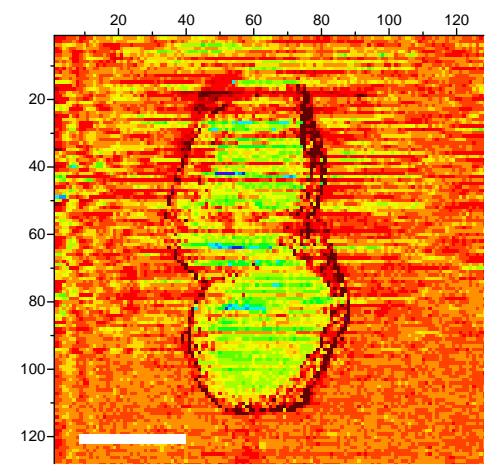

Figure SI03.d 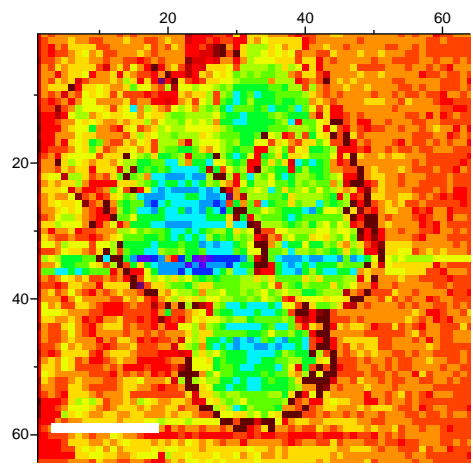 Figure SI03.e 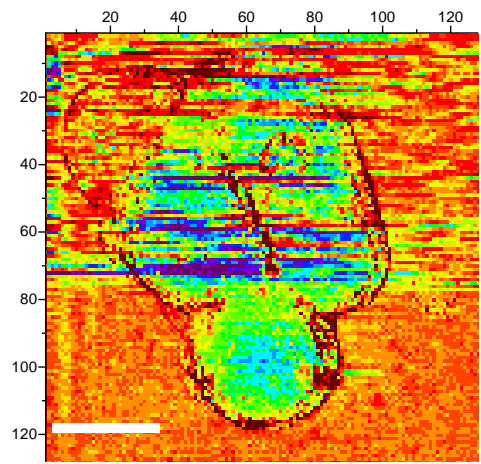

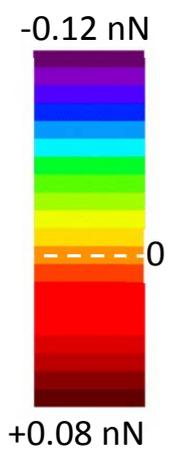

−0.12 nN

0

+0.08 nN

SI6

Figure SI04.a

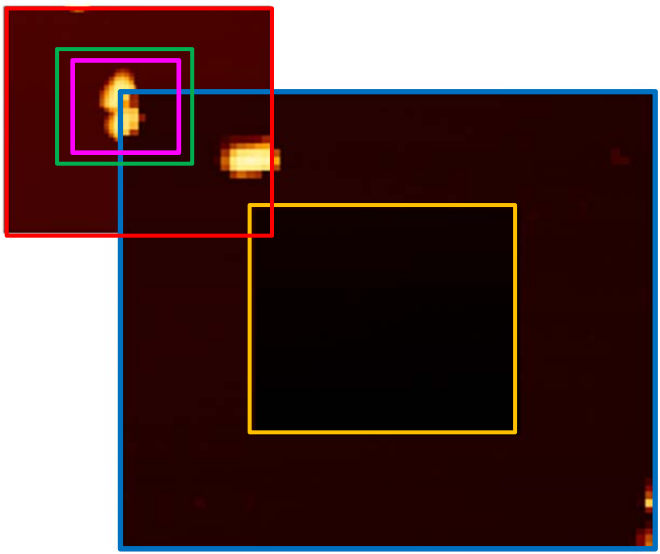

Figure SI04.b

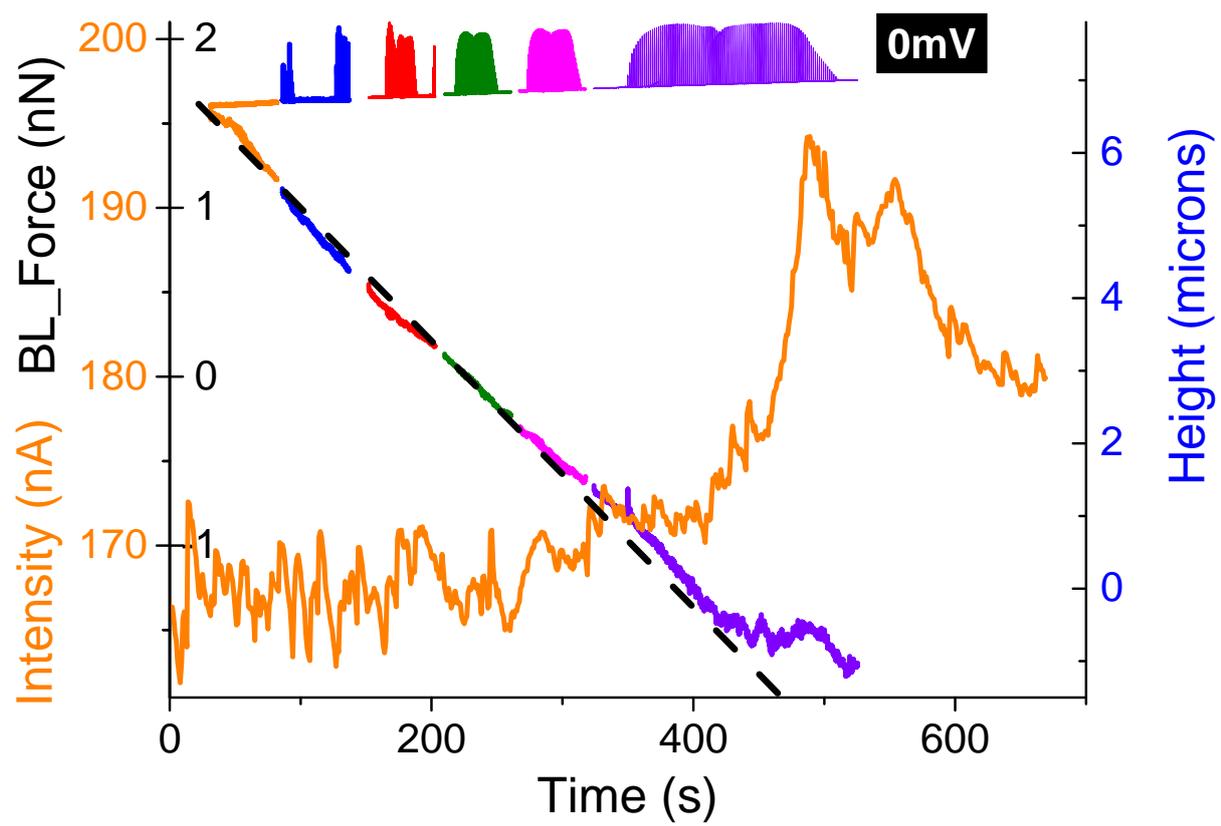

SI7

Figure SI05. a

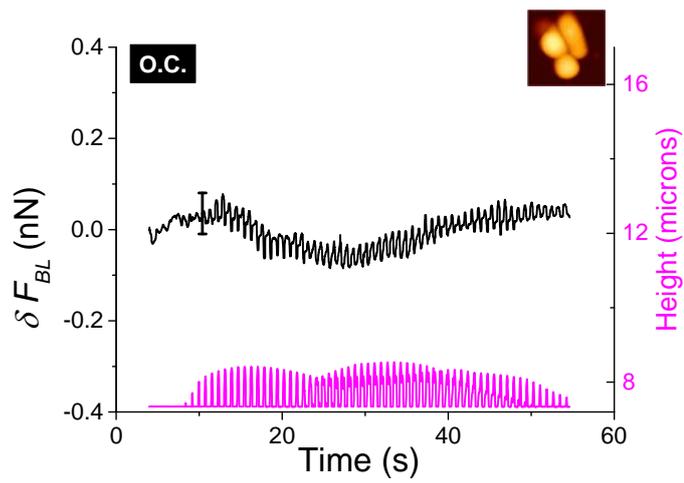

Figure SI05.b

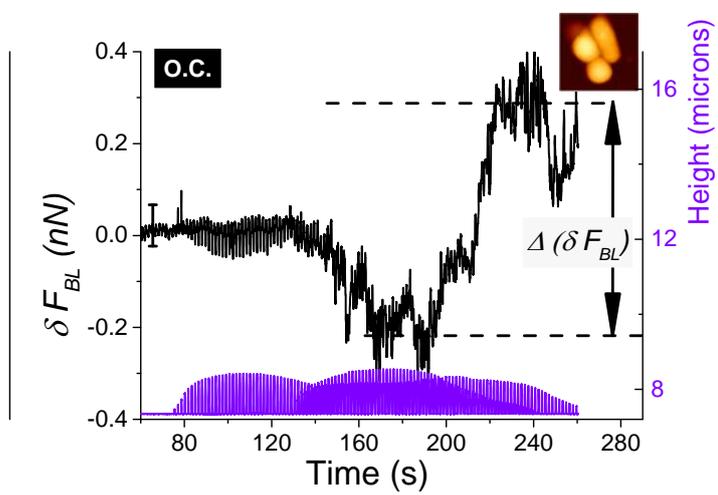

Figure SI06.a

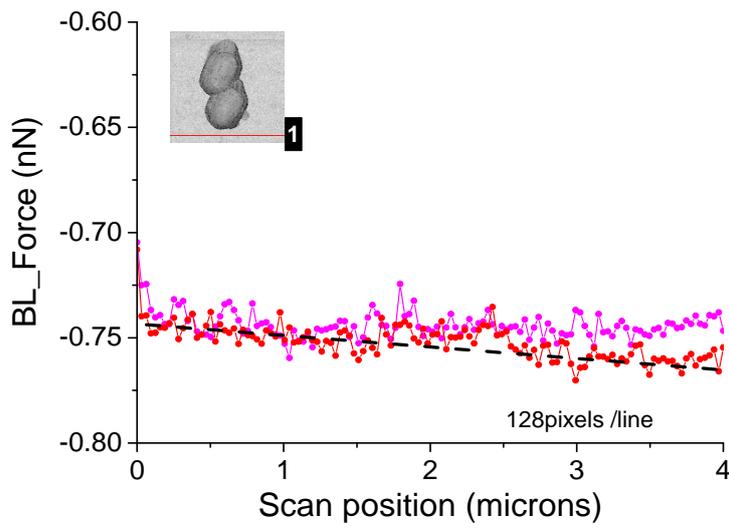

Figure SI06.b

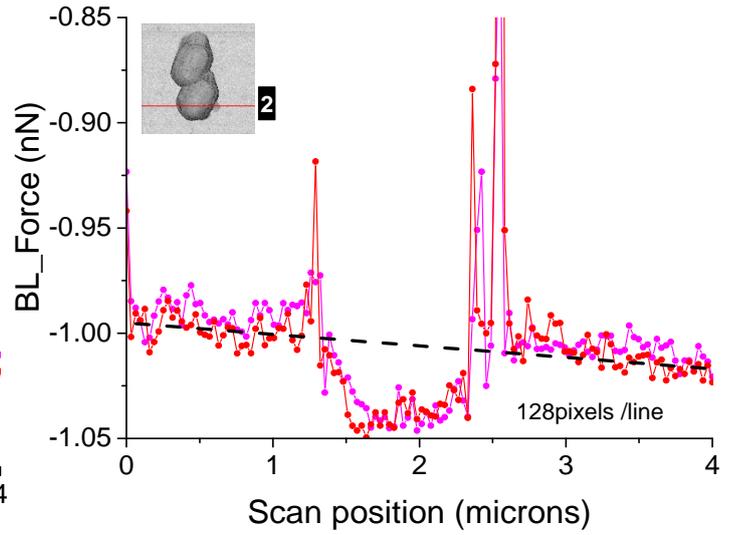

Figure SI06.c

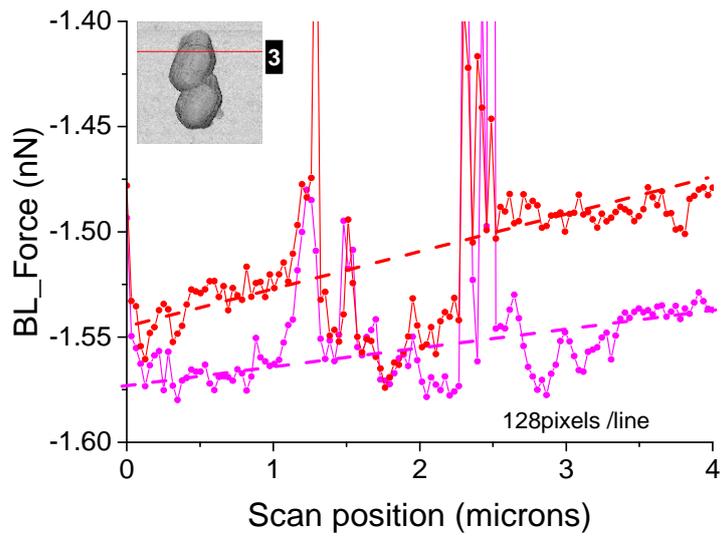

Figure SI06.d

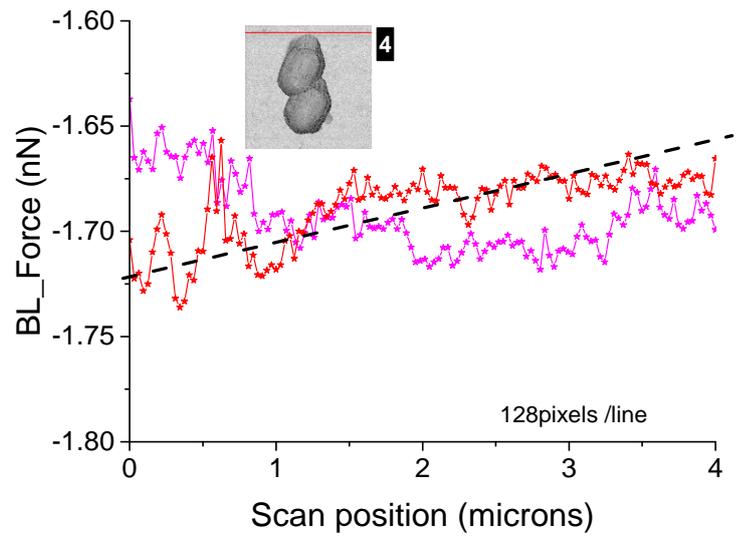

SI9

Figure SI07.a

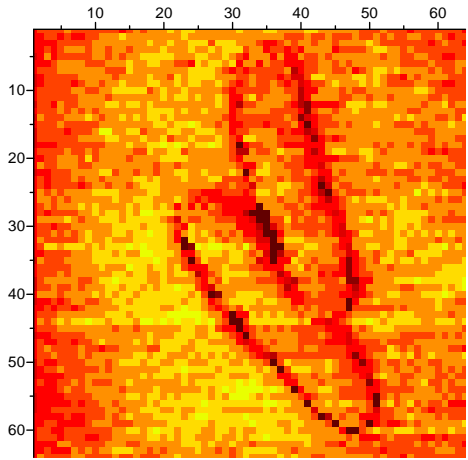

Figure SI07.b

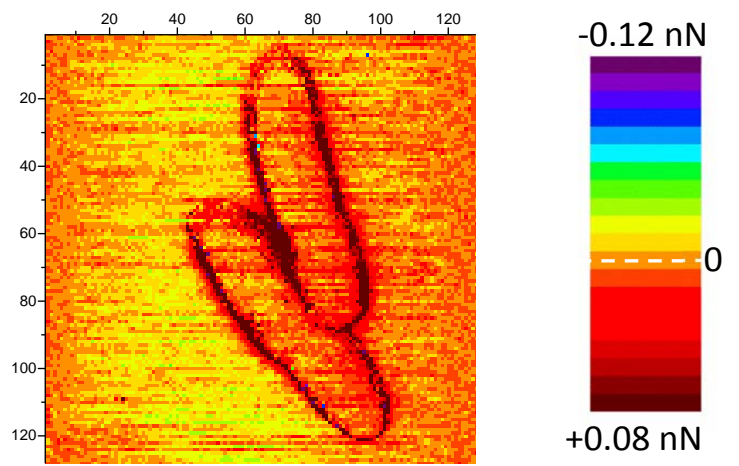

Figure SI07.c

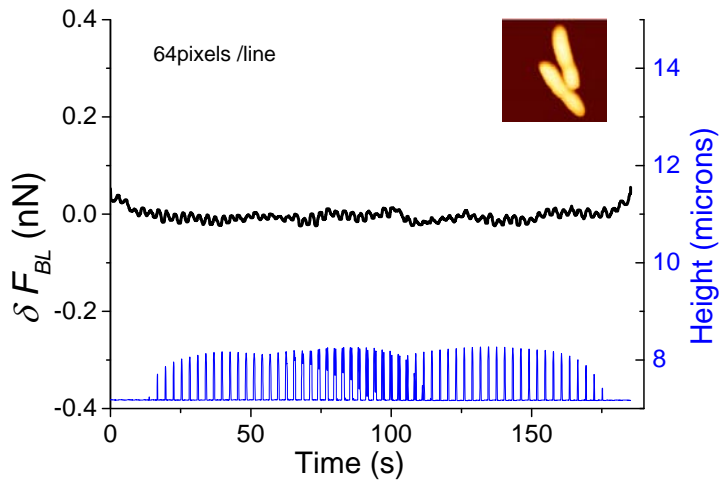

Figure SI07.d

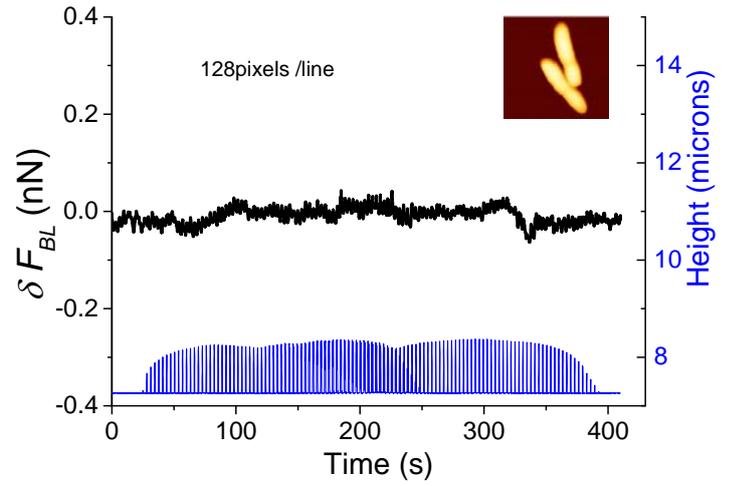

SI10